\documentstyle[epsfig]{mn2e}
\newif\ifAMStwofonts                                                                                       

\def\gsimeq{{_>\atop^{\sim}}}

\def\eso36{\hbox{ESO~3.6-m}}                          
\begin{document}

\title[{An XMM-{\it Newton} hard X-ray survey of ULIRGs}]
{AN XMM-{\it Newton} HARD X-RAY SURVEY OF ULTRA-LUMINOUS INFRARED GALAXIES
\thanks{Based on observations obtained with {\it XMM-{\it Newton}}, an
ESA science missions with instruments
and contributions funded by ESA Member States and the USA (NASA).}
} 
 \author[A. Franceschini, V. Braito, M. Persic, et al.] 
{\parbox[]{6.5in}{A.\,Franceschini$^1$\thanks{e-mail: franceschinini@pd.astro.it},  
V.\,Braito$^{1,2}$, M.\,Persic$^{3}$, R.\,Della Ceca$^{4}$,    
L.\,Bassani$^{5}$, M.\,Cappi$^{5}$, G.\,Malaguti$^{5}$,   
G.G.C.\,Palumbo$^{6}$, G.\,Risaliti$^{7,8}$,  
 M.\,Salvati$^{7}$ and P.\,Severgnini$^{3}$ } 
\\ 
 $^1$Dipartimento di Astronomia, Universit\`a di  Padova, Vicolo Osservatorio 2, 
  I-35122, Italy  
\\ 
$^2$ INAF$-$Osservatorio Astronomico di Padova, Italy   
\\ 
$^3$INAF$-$Osservatorio Astronomico di Trieste,  Italy  
\\ 
$^4$INAF$-$Osservatorio Astronomico di Brera, Milano, Italy  
\\ 
$^5$IASF $-$ CNR, Bologna, Italy 
\\ 
$^6$Dipartimento di Astronomia, Universit\`a di Bologna,  Italy  
\\ 
$^7$INAF$-$Osservatorio Astrofisico di Arcetri, Firenze, Italy.\\  
$^8$ Harvard-Smithsonian Center for Astrophysics, Cambridge, USA
}

\maketitle 
 
\date{Accepted 2002 December 15. Received 2001 February 21;  
in original form 2000 November 13} 
 
\pagerange{\pageref{firstpage}--\pageref{lastpage}}  
 

\label{firstpage} 
 
\begin{abstract} 
XMM-{\sl Newton} observations of 10 Ultra-Luminous Infrared Galaxies 
(ULIRGs) from a 200 ksec mini-survey program are reported. The
aim is to investigate in hard X-rays a complete ULIRG 
sample selected from the bright IRAS 60 $\mu$m catalogue.
All sources are detected in X-rays, 5 of which for the first time. These observations confirm that ULIRGs are 
intrinsically faint X-rays sources, their observed X-ray luminosities being typically 
$L_{2-10 keV}\leq 10^{42-43}\ erg/s$, whereas their bolometric (mostly IR) 
luminosities are $L_{bol}>10^{45}\ erg/s$.
In all sources we find evidence for thermal emission from hot plasma with a
rather costant temperature $kT\simeq 0.7$ keV, dominating the X-ray spectra below
1 keV, and likely associated with a nuclear or circumnuclear starburst. 
This thermal emission appears uncorrelated with the far-IR luminosity, suggesting
that, in addition to the ongoing rate of star formation, other parameters may also 
affect it.    The soft X-ray emission appears to be
extended on a scale of $\sim$30 kpc for Mkn 231 and IRAS 19254-7245, 
possible evidence of galactic superwinds. In these two sources, in IRAS 20551-4250 
and IRAS 23128-5919 we find evidence for the presence of hidden AGNs, while a 
minor AGN contribution may be suspected also in IRAS 20100-4156. In particular, 
we have detected a strong (EW$\sim 2$ keV) Fe-K line at 6.4 keV in the spectrum of 
IRAS19254-7245 and a weaker one in Mkn 231, suggestive of deeply buried AGNs.
For the other sources, the X-ray luminosities and spectral shapes are consistent 
with hot thermal plasma and X-ray binary emissions of mainly starburst origin. We find 
that the 2-10 keV luminosities in these sources, most likely due to high-mass X-ray 
binaries, are correlated with L$_{FIR}$: both luminosities are good indicators of the 
current global star formation rate in the galaxy.
The composite nature of ULIRGs is then confirmed, with hints for a predominance 
of the starburst over the AGN phenomenon in these objects even when observed in 
hard X-rays.

\end{abstract} 
 

\section{Introduction} 

Ultra-Luminous InfraRed Galaxies (ULIRGs, sources with bolometric luminosity  
L$_{IR}> 10^{12}$L$_\odot$ mostly emitted in the IR) have received much attention  
since their discovery during the IRAS survey follow-up observations (Sanders et al.  
1988; see a review in Sanders \& Mirabel 1996). 
The main reason for this interest was that, together with optical quasars, these 
are by far the most luminous objects in the local universe. 

The relevance of this class of sources in the cosmological context has been  
further emphasized by recent findings of cosmological surveys at long 
wavelengths. Observations at IR and sub-millimeter wavelengths have discovered 
that luminous and ultra-luminous IR galaxies, which are rare in the local 
universe (Soifer et al. 1987), are detected instead in large numbers 
in deep IR surveys, and are a fundamental constituent of the high-redshift 
galaxy population (e.g. Smail et al. 1997; Genzel \& Cesarsky 2000; 
Franceschini et al. 2001). The number and luminosities of these sources imply 
that an important fraction of stars in present-day galaxies, or alternatively 
of the degenerate baryons contained in nuclear supermassive Black Holes, 
have formed during such IR-luminous dust-extinguished evolutionary phases
in the past. 
It has been argued that luminous and ultra-luminous IR galaxies at high 
redshifts could trace events of star-formation which may be at the origin of 
massive elliptical and S0 galaxies 
(e.g. Franceschini et al. 1994; Lilly et al. 1999). Support for this view 
has recently come from high-resolution spectroscopy of ULIRG mergers (Genzel 
et al. 2001), whose dynamical properties have proven that they indeed 
are "ellipticals in formation".

This evidence came along with the discovery by COBE of a bright diffuse radiation, 
the Cosmic IR Background (CIRB, see Hauser et al. 1998; Lagache et al. 2000), 
appearently containing a large fraction (up to $\sim 70$\%) of the  
total extragalactic background energy density from radio to X-rays, i.e.
a major part of the photon energy released by cosmic sources at any redshifts. 
As discussed by various authors (Elbaz et al. 2002; Smail et al. 2002), 
there is precise evidence that this background radiation  
is indeed produced by sources similar in all respects to luminous and  
ultra-luminous IR galaxies at $z\geq 0.5$. 
These recent facts justify the growing interest in local ULIRGs as possible
clues to their supposed high-z counterparts. 
 
However, about 15 years after their discovery, 
the nature of these sources still remains rather enigmatic.
Large gas and dust column densities in the galaxy cores, responsible for  
the IR emission, prevent a direct observation of the primary energy source and the 
IR spectral shapes are highly degenerate with respect to the illuminating spectrum.
It is widely accepted that both starburst and AGN activity may be  
responsible for the observed luminosities (Genzel et al. 1998; Veilleux, Kim, 
Sanders 1999; Risaliti et al. 2000; Bassani et al. 2003, in prep.).
If indeed ULIRGs are "forming spheroids", an AGN/starburst association 
is naturally supported by the evidence that all spheroidal galaxies host quasar 
relics in the form of supermassive BH's. The same radial
inflow of gas produced by the merger is likely to fuel both the starburst and,
at some stages, the AGN.

So far conflicting evidence has been reported about the relative contributions  
of the two energy sources in ULIRGs, stellar and gravitational. 
Important progress has been achieved with mid- 
and far-IR spectroscopy, probing the inner optically-thick nuclei. Using 
diagnostics based on coronal line intensities and PAH line-to-continuum ratios,  
Lutz et al. (1996), Genzel et al. (1998) and Rigopoulou et al. (1999) have argued 
that the majority of ULIRGs are powered by star-formation. 

On the other end, the presence of energetically important obscured AGNs in ULIRGs 
has been revealed by optical and near-IR spectroscopy, often detecting  
Seyfert-like nuclear emission line spectra (Sanders et al. 1988; 
Veilleux, Sanders \& Kim, 1997) as well as evidence for completely buried 
AGNs (Soifer et al. 2001; Imanishi, Dudley \& Maloney 2001). 
 
In principle hard X-ray observations offer an additional important tool to investigate  
the presence of hidden AGNs, providing quantitative estimates of their  
contribution to the bolometric luminosity. This diagnostics relies on the 
fundamentally different spectra and luminosity regimes between starbursts  
and AGNs in hard X-rays. Hard continuum emission and prominient the Fe-K$\alpha$ line(s) are
distinguishing features of buried AGNs, which can penetrate large gas column densities. 
Some of the brightest nearby ULIRGs, classified on the basis of IR spectroscopy 
as pure starburst, show spectral properties typical of obscured AGNs when  
observed in hard X-rays (e.g. NGC 6240, Iwasawa et al. 1999; Vignati et al.1999). 
In practice, so far the utilization of this diagnostics has been limited by the 
X-ray faintness of most ULIRGs (see previous unsuccessful detection attempts with 
ROSAT, ASCA and BeppoSAX; Risaliti et al. 2000). 

The unique large collective area and hard X-ray response of XMM-{\sl Newton} have been
used here to deeply survey for the first time a representative 
sample of 10 IRAS-selected ULIRGs for which high-quality mid-IR and optical 
spectroscopy data are available. Most of these ULIRGs were previously classified as  
starbursts (Lutz, Veilleux \& Genzel 1999). 
The purpose of the present observations
is not only to identify AGN signatures in hard X-rays, but also 
to separately quantify the contributions of the buried AGN from the starburst emission to
the X-ray spectra, and to compare them with independent estimates based on optical and 
IR observations. 
This paper summarizes the main X-ray properties of the sample, five sources of
which are detected in X-rays for the first time.
XMM-{\sl Newton} observations for two sources of the sample of particular interest, the
Superantennae and Mkn 231, are discussed in separate papers (Braito et al. 2003a and
2003b).

Section 2 summarizes the properties of our ULIRG sample.
Section 3 details the XMM-{\sl Newton} observations and data analysis.
The resulting hard X-ray properties are reported in Section 4, while
Section 5 discusses plausible origins of various X-ray components and
compares them with data at other wavelengths. Section 6 contains our conclusions. 
$H_0=50$ $Km~s^{-1}$ $Mpc^{-1}$ has been assumed throught.

\begin{figure*}
\begin{tabular}{cc}
\epsfig{file=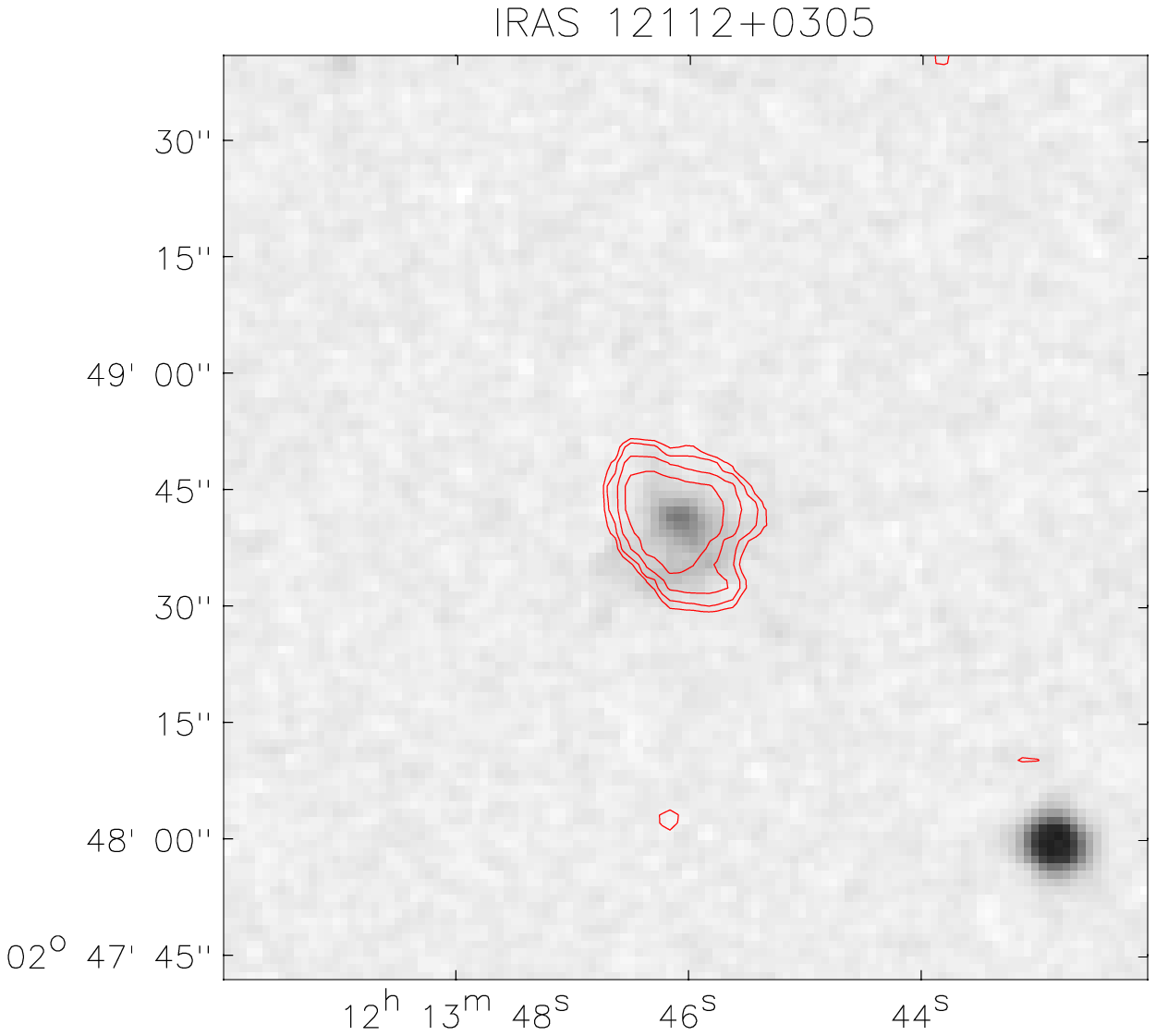,width=0.7\textwidth}&\hskip -4truecm\epsfig{file=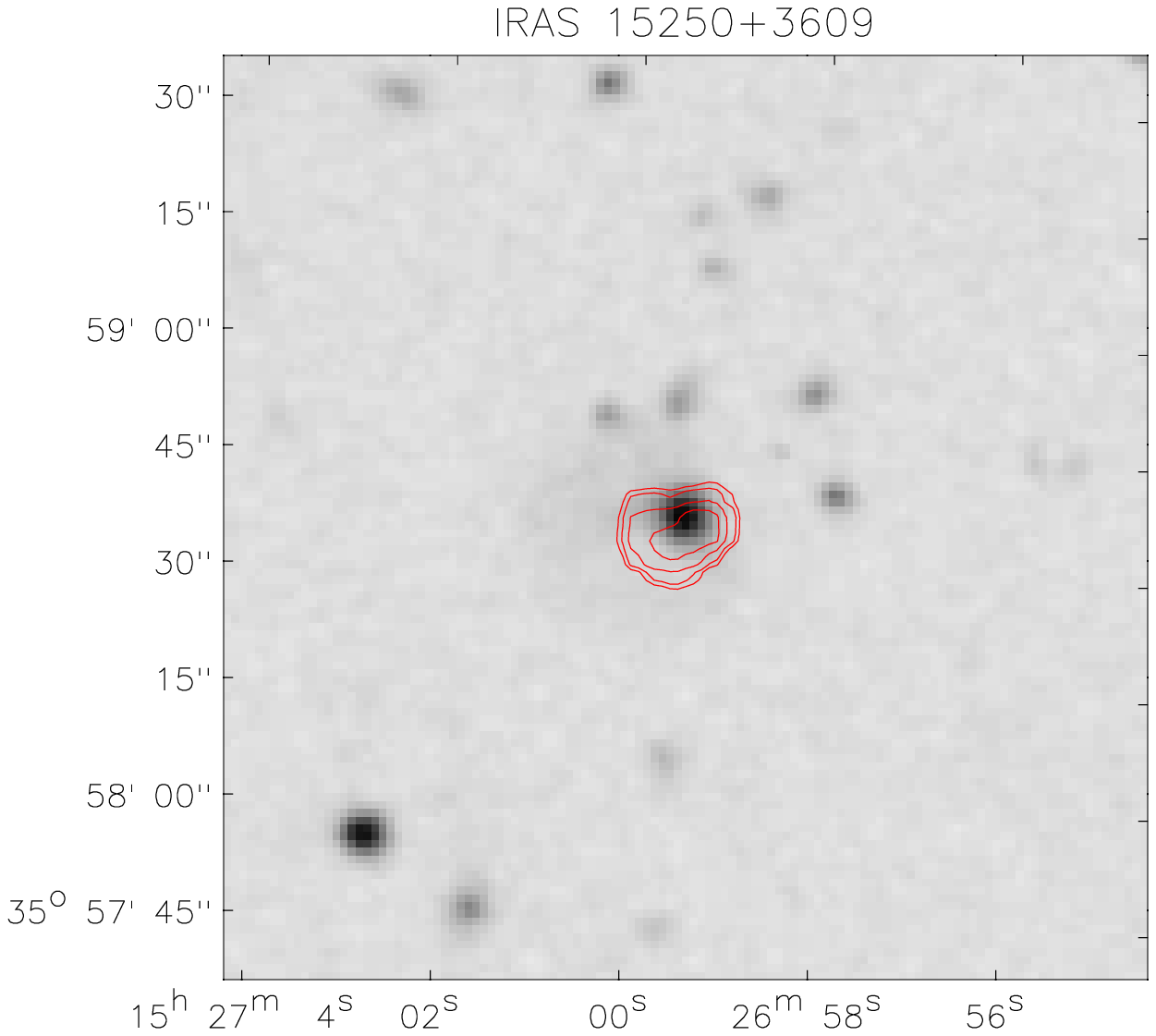,width=0.7\textwidth}\\
\epsfig{file=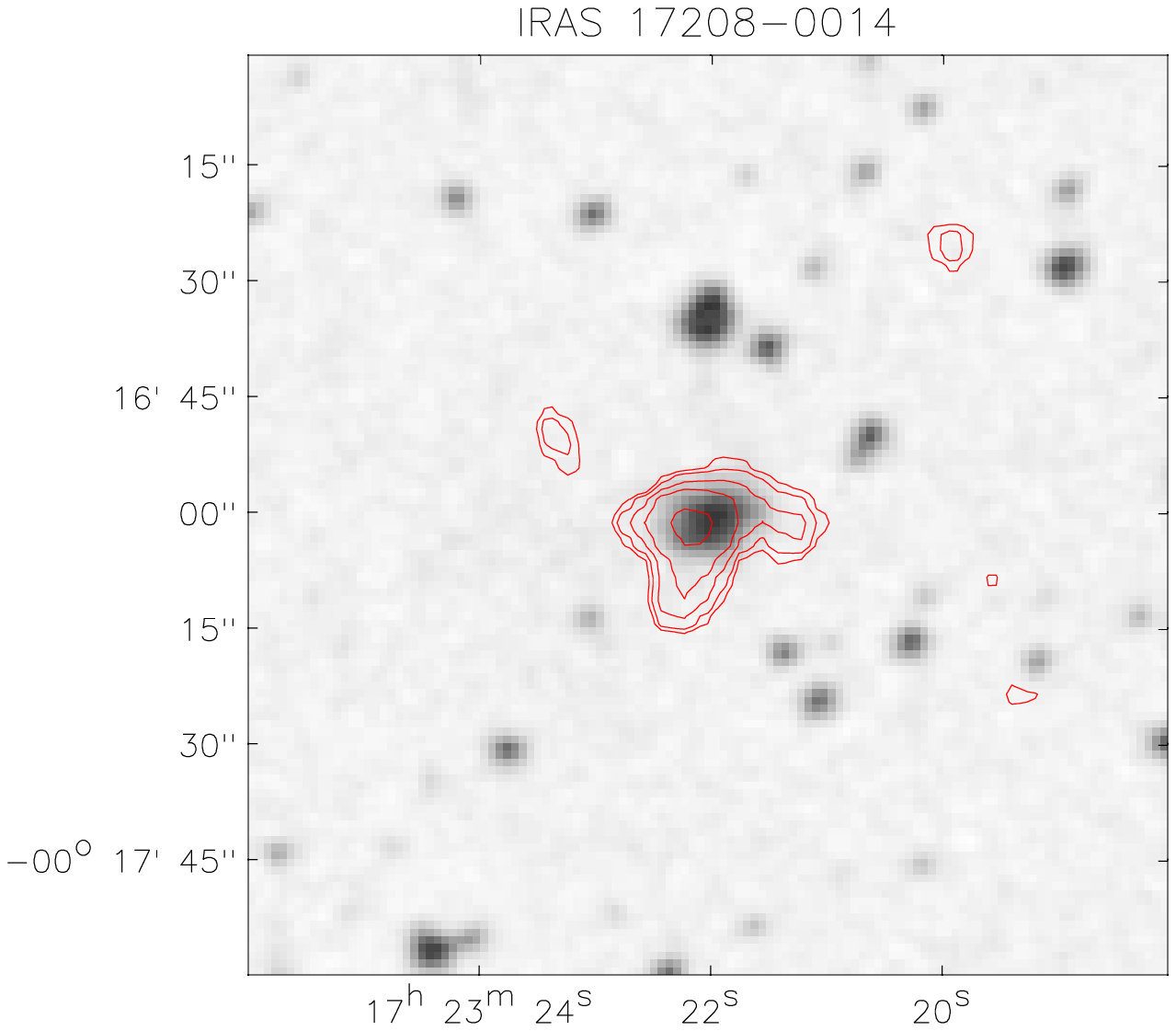,width=0.7\textwidth}&\hskip -4truecm\epsfig{file=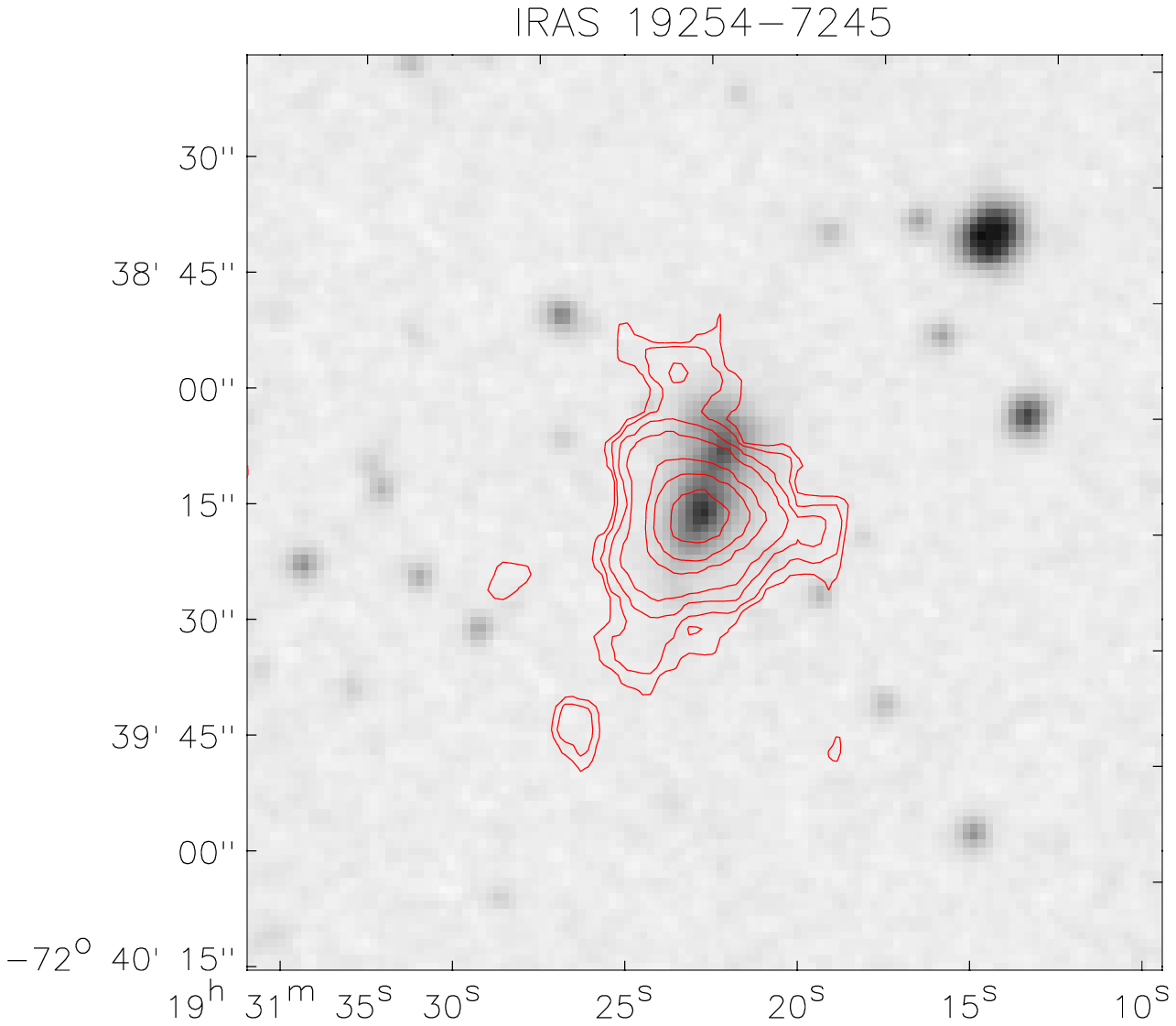,width=0.7\textwidth}\\
\end{tabular}
\caption{DSS2 images ($2'\times 2'$) of IRAS 12112$+$0305, IRAS$15250+3609$, 
IRAS$17208-0014$ and IRAS$19254-7245$ (from top left to bottom right). Countours 
of the X-ray (0.2$-$10 keV)  emission have been overlayed. The countours displayed 
correspond to 4$\sigma$, 5$\sigma$, 7$\sigma$, 10$\sigma$, 20$\sigma$, 30$\sigma$, 
50$\sigma$ above the background. The countours have been derived using the MOS2 data, 
since this detector has the best point spead function (see Ehele et al. 2001).}
\label{fig1}
\end{figure*}

\begin{figure*}
\begin{tabular}{cc}
\epsfig{file=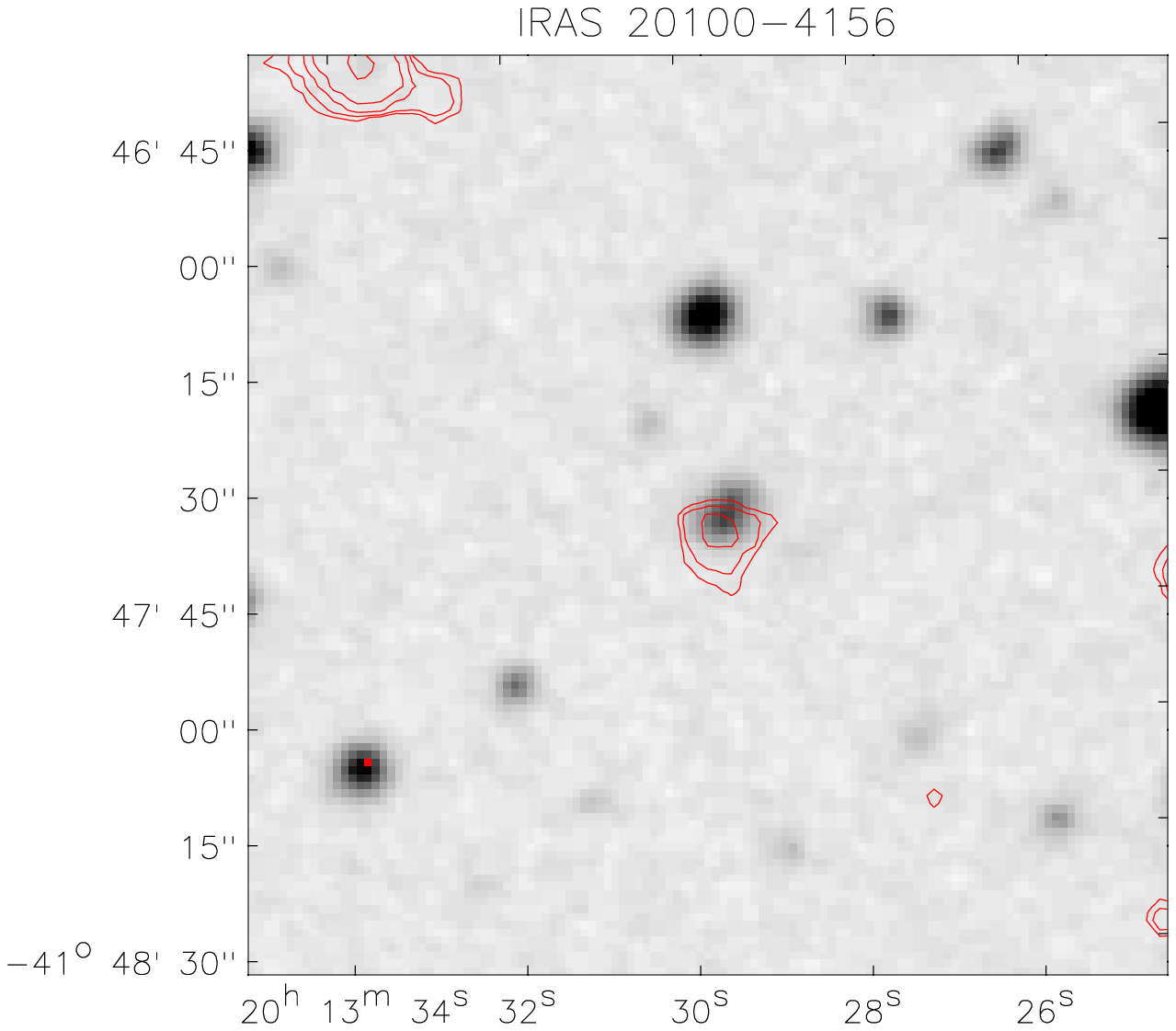,width=0.7\textwidth}&\hskip -4truecm\epsfig{file= 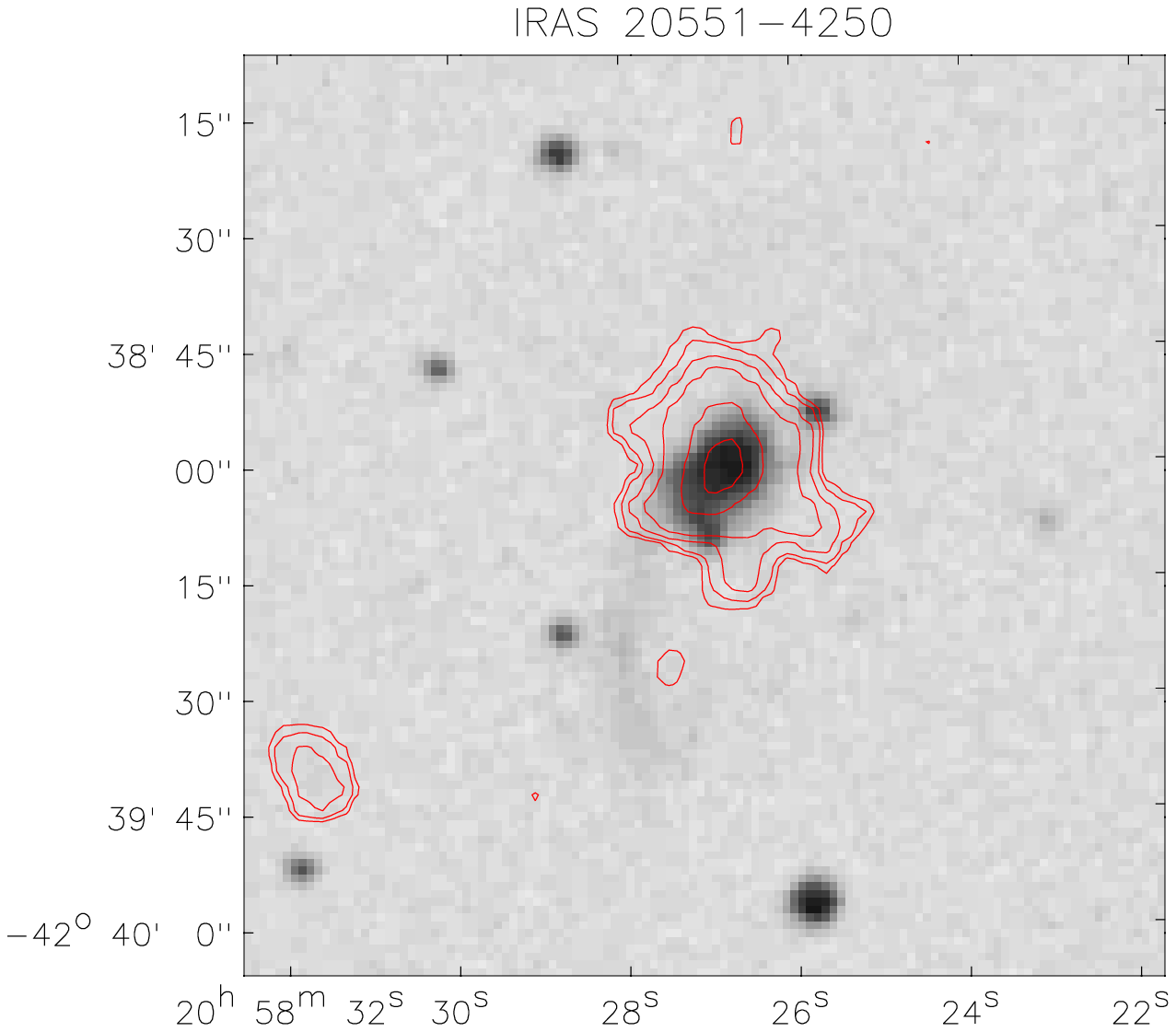,width=0.7\textwidth}\\
\epsfig{file=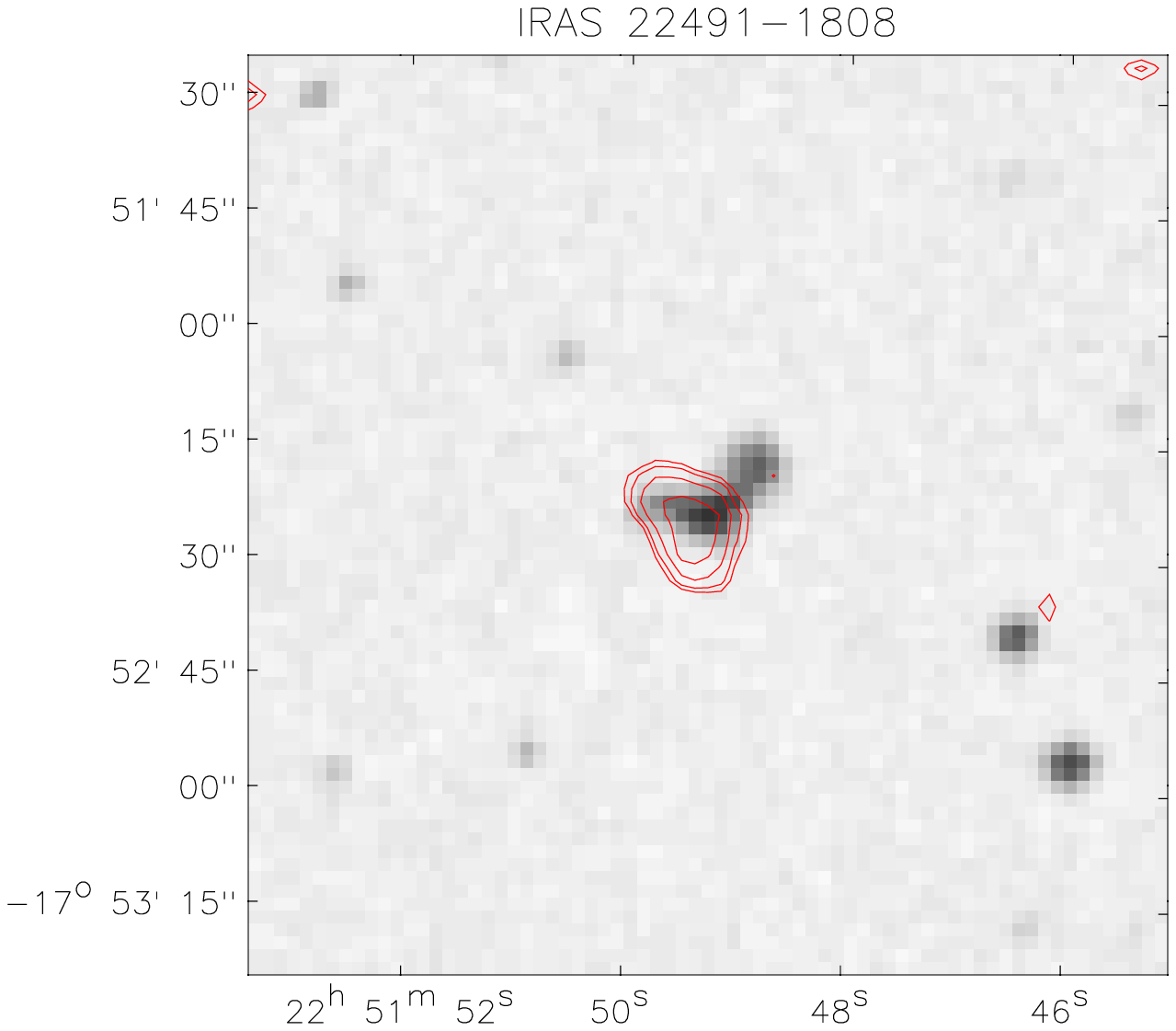,width=0.7\textwidth}&\hskip -4truecm\epsfig{file=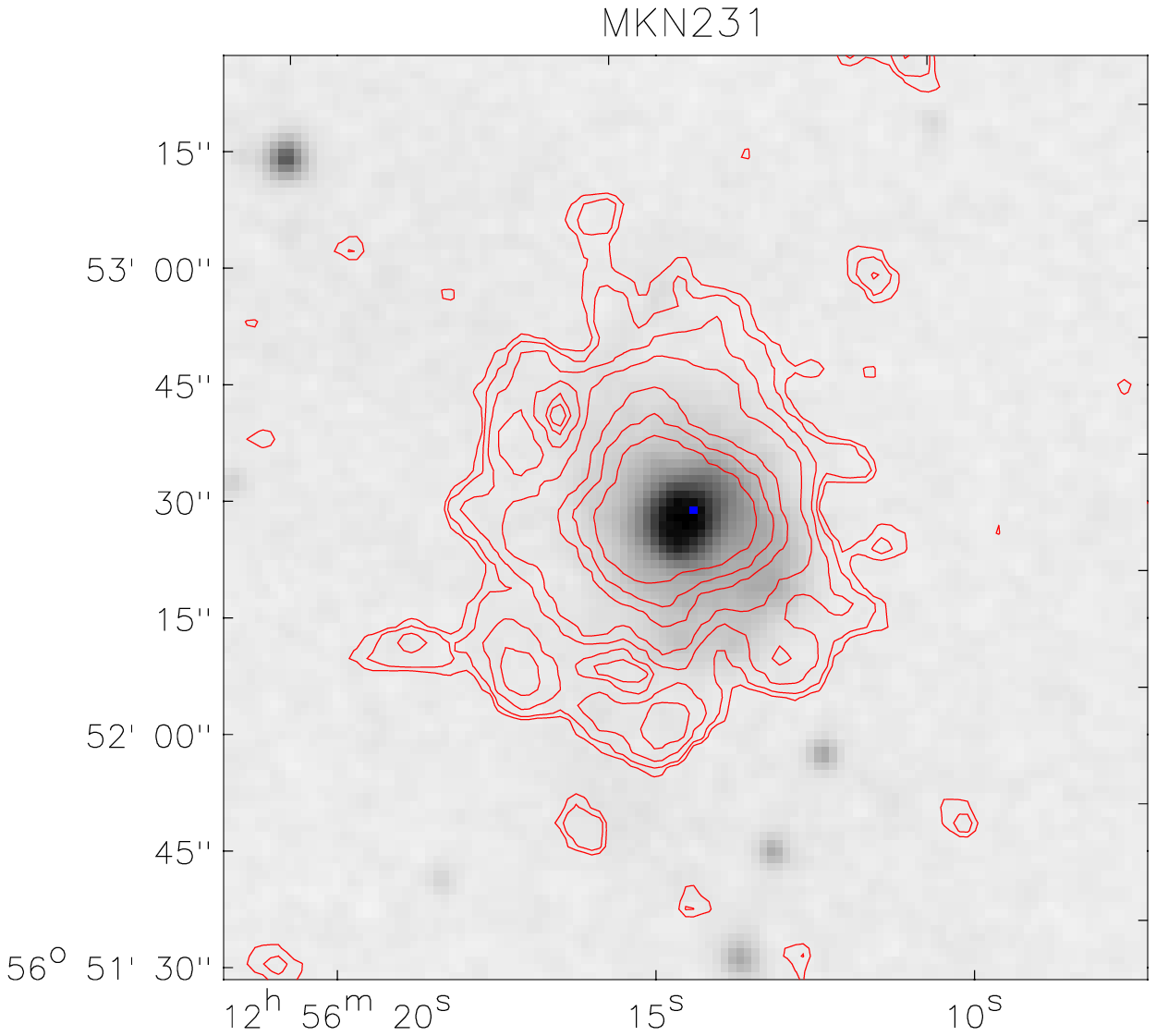,width=0.7\textwidth}\\
\end{tabular}
\caption{DSS2 images ($2'\times 2'$) of IRAS 20100$-$4156, IRAS$20551-4250$,  
IRAS$22491-1808$ and MKN231 (from top left to bottom right). 
Countours of the X-ray (0.2$-$10 keV)  emission have been overlayed. The countours 
displayed (from the MOS2 data) correspond to 4$\sigma$, 5$\sigma$, 7$\sigma$, 10$\sigma$, 20$\sigma$, 
30$\sigma$, 50$\sigma$ above the background. See also caption to Fig.1 }
\label{fig2}
\end{figure*}

\begin{figure*}
\begin{tabular}{cc}
\epsfig{file=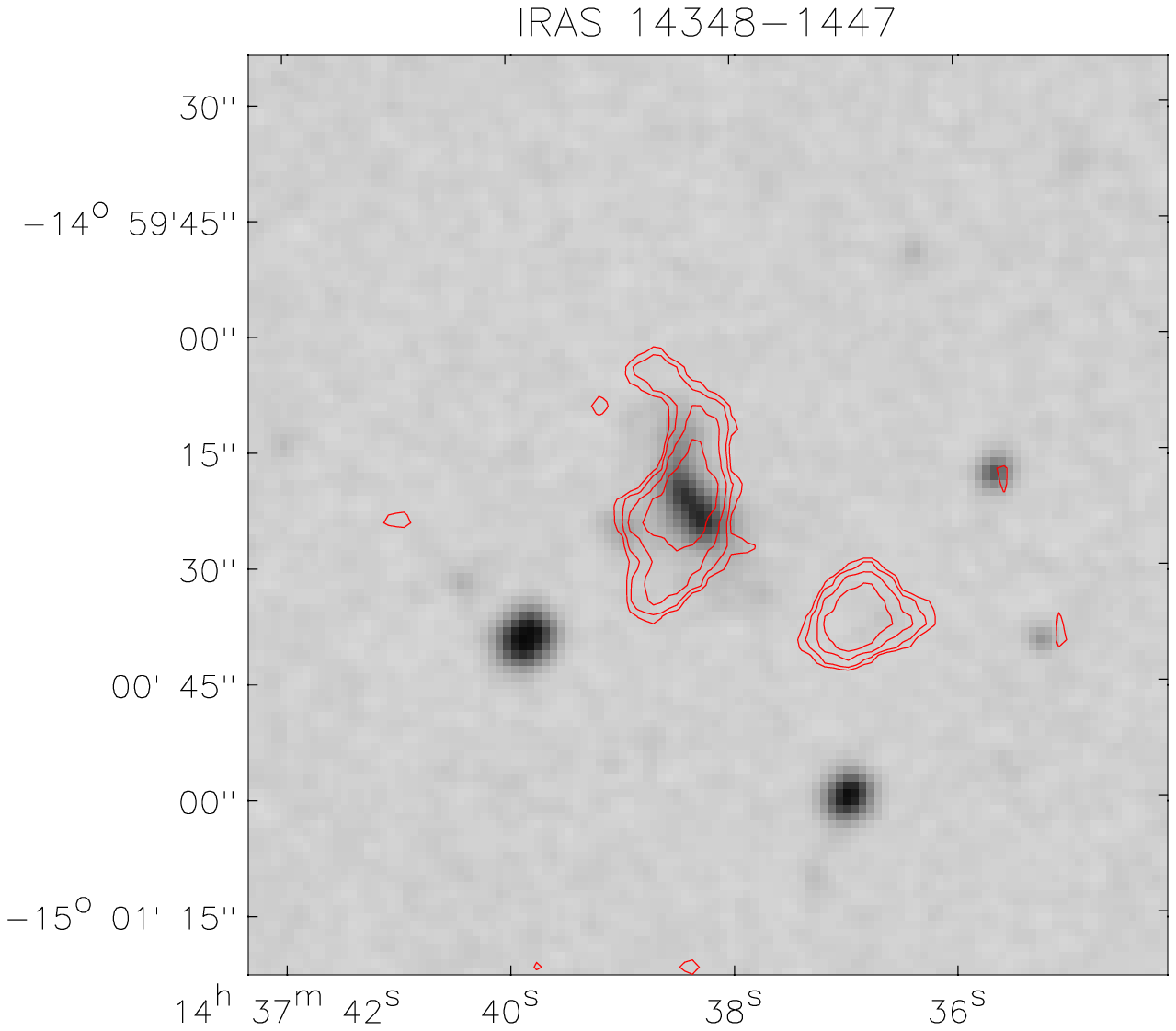,width=0.7\textwidth}&\hskip -4truecm
\epsfig{file=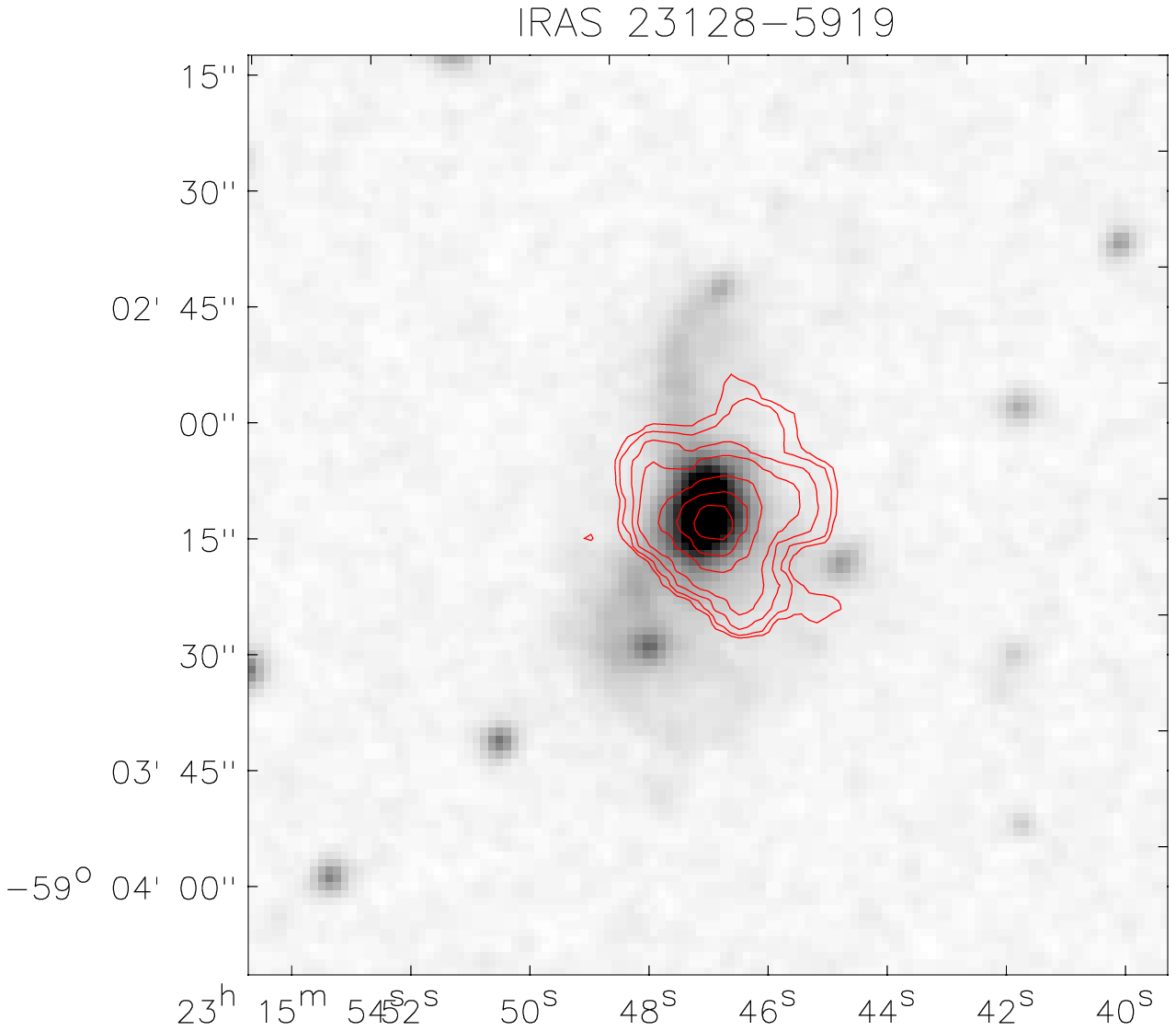,width=0.7\textwidth}\\
\end{tabular}
\caption{DSS2 images and XMM-Newton contours ($2'\times 2'$) of IRAS 14348$-$1447
and IRAS 23128-5919. Countours of the X-ray 
(0.2$-$10 keV)  emission have been overlayed. The countours displayed 
(from the MOS2 data) correspond to 
4$\sigma$, 5$\sigma$, 7$\sigma$, 10$\sigma$, 20$\sigma$, 30$\sigma$, 50$\sigma$ above 
the background. See also caption to Fig.1.}
\label{fig3}
\end{figure*}

\section{The ULIRG SAMPLE} \label{vbraito-E1_sec:XMM}  
 
The sample of 10 ULIRGs discussed in this paper has been selected from the list of 
IRAS ULIRGs observed with ISO by Genzel el al. (1998). The sample is flux-limited 
at $60\mu$m and complete to $S_{60\mu}\geq 5.4$ Jy. It includes sources with  
$L\geq 10^{12}$ L$_{\odot}$ in the total 8--1000 $\mu$m band, whose
far-IR selection makes it unbiased with respect to absorption and
representative of the most luminous galaxies in the local universe.
High quality IR/optical spectroscopic data are available for all the   
sample sources (Lutz et al. 1999; Veilleux et al. 1999). 

Of the original Genzel et al. sample of 15 ULIRGs, 4 have been observed with 
XMM-{\it Newton} by different teams, while IRAS 23060+0505 has been deeply observed
with ASCA (Brandt et al. 1997) and BeppoSAX.
Source names, sky positions (which are also the coordinates of the XMM-EPIC observations),
IR (8-1000 $\mu$m) luminosities, redshifts, as well as the XMM-EPIC net exposure 
times and observation dates, of the 10 ULIRGs presented here, are reported in Table 1.  
Optical and IR classifications of the ULIRGs presented here are summarized 
in Table 4. 

All 10 objects have been observed with XMM-{\it Newton} with $\sim$20 ksec 
exposure each. A fraction of the observations had to be excluded from the analysis
because of background flares (this problem was severe for the source
IRAS 23128-5919, see Table 1).

\section{XMM-{\it Newton} EPIC Data Preparation and Analysis}  
\label{fauthor-E1_sec:an}  
  
The XMM-{\it Newton} observations presented here have been performed between March 2001 and
November 2002 with the EPIC (European Photon Imaging Camera: Str\"uder et al. 2001  and 
Turner et al. 2001) cameras  operating in full-frame mode.  
Data have been processed using the Science Analysis Software (SAS version 5.3),
and have been analyzed using standard software packages (FTOOLS 5.0, XSPEC 11.0).
The latest calibration files  released by the EPIC team have been used.  

Event files produced from the pipeline have been filtered from high-background time 
intervals, and only events corresponding to pattern 0-12 for MOS and pattern 0-4 for 
PN have been used (see the XMM-{\it Newton} Users' Handbook, Ehle et  al. 2001). 
We have then generated our own response matrices (that include the correction 
for the effective area) using the SAS tasks {\it arfgen} and {\it rmfgen}.
The screening process from high background time intervals yielded net exposures 
between 15.7 ksec  and 21.7 ksec (net exposures for PN camera are reported in  
Table 1), with the exception of IRAS 23128-5919 (9.4 ksec).    
  
All the 10 ULIRGs have been detected in the MOS1, MOS2 and PN detectors   
with signal to noise ratios S/N$>$3.   
Close-up images of the sources are reported in Figs. \ref{fig1}, \ref{fig2},
\ref{fig3} as overlays of the (0.2--10 keV) X-ray contours on top of the POSS 
red-band images.
For all the ULIRGs the bulk of the X-ray  emission is positionally concident 
(within the XMM-{\it Newton} positional error) with the optical-IR core.
By accounting in detail for the EPIC PSF, we have found evidence for extended emission
only in Mkn 231 and IRAS19254-7245, while the surface brightness distributions
for all other sources are consistent with being unresolved by EPIC.

Except for Mkn 231, the most extended source
(90\% counts in $\sim$ 40 arcsec in the soft X-ray band E$<$ 2 keV), the 
source spectra were extracted from circular regions of $\sim 20$ arcsec radius.   
Background spectra have been extracted from source-free circular regions (with 
typically $\sim 1^\prime$ radius) close to the target. 

In order to improve the statistics, MOS1 and MOS2 data have been combined.  
Both MOS and PN spectra were then automatically rebinned in order to have a 
$\geq 20$  counts in each energy channel. 
The combined MOS data have been fitted simultaneously with the
PN data, by keeping the relative normalizations free.

\begin{table*}  
  \label{tab1}  
  \leavevmode  
  \footnotesize  
  \caption{The source sample and summary of the XMM-{\it Newton} observations}  
  \begin{tabular}[h]{lccccccl}  
  \hline \\[-5pt]  
    Name                 & Coordinates               & z     & L$_{IR}$ & PN Net exp. time& Obs. Date &Sequence&
    Filter \\[+5pt]  
                         & (J2000)                   & &[$10^{12}$ L$_\odot$]& [ksec] & [month/year] & \\   
\hline  
IRAS 12112$+$0305        & 12 13 46.03 $+$02 48 41.5 & 0.072 & 3.8      & 20.9     & 2001 Dec 30 & 0081340801&Medium (PN) -Thin       \\  
MKN231$^{a}$             & 12 56 14.16 $+$56 52 24.9 & 0.042 & 6.7      & 19.8     & 2001 Jun 06& 0081340201 &Medium  \\  
IRAS 14348$-$1447        & 14 37 38.30 $-$15 00 23.0 & 0.082 & 4.0       & 17.5     & 2002 Jul 29 & 0081341401 &Medium       \\  
IRAS 15250$+$3609        & 15 26 59.43 $+$35 58 37.4 & 0.055 & 2.0      & 17.8     & 2002 Feb 22 & 0081341101 &Thin	     \\  
IRAS 17208$-$0014        & 17 23 22.03 $-$00 17 00.3 & 0.043 & 4.7      & 15.7     & 2002 Feb 19 &0081340601 &Medium 	\\  
IRAS 19254$-$7245$^{b}$  & 19 31 21.46 $-$72 39 21.6 & 0.062 & 2.2      & 18.4     & 2001 Mar 21&0081341001& Thin	\\  
IRAS 20100$-$4156        & 20 13 29.75 $-$41 47 34.0 & 0.129 & 7.4      & 18.1     & 2001 Apr 21& 0081340501 & Medium \\  
IRAS 20551$-$4250        & 20 58 27.05 $-$42 39 06.8 & 0.043 & 2.1      & 16.1     & 2001 Apr 21& 0081340401 & Thin   \\   
IRAS 22491$-$1808        & 22 51 49.26 $-$17 52 24.0 & 0.078 & 2.7      & 21.7     & 2001 May 24&  0081340901 & Medium \\  
IRAS 23128$-$5919        & 23 15 47.00 $-$59 03 17.0 & 0.044 & 2.0      &  9.4     & 2002 Nov 19&  0081340301 & Medium \\  
\hline \\ 
 \end{tabular} 
\begin{flushleft}
{\em  Note: $^a$: XMM{-\sl Newton} observations reported in Braito et al. 2003b; \\
       ~~~~~~ $^b$: XMM{-\sl Newton} observations reported in Braito et al. 2003a} 
 \end{flushleft}
 \end{table*}

\section{Results}  \label{results}

We report in Table \ref{tab3} the measured fluxes and luminosities 
of the sources in various X-ray bands. Our XMM-{\it Newton} observations show 
that ULIRGs are rather faint in X-rays if compared with AGNs of similar bolometric 
(IR) luminosities. Their observed X-ray luminosities are close to or lower than 
$L_{2-10 keV}=10^{42}\ erg/s$, while
their IR bolometric output is always $L_{bol}>10^{45}\ erg/s$.

The level of detail for our spectral analysis is different for each source
depending on the qualities of the XMM spectra, ranging from fairly
detailed for stronger sources to only rather coarse spectral fits for the faintest. 
The two brightest sources,
IRAS19254-7245 and Mkn 231, have data of sufficiently high quality to guarantee
investigations of both the continuum emission and the Fe-K 6.4 keV line,
detected only in these two sources (see Braito et al. 2003a,b).

At energies above 1 keV, the X-ray spectra appear to be dominated by a relatively
flat power-law (PL) component in most of the sources. 
We tried to fit these overall 0.2-10 keV spectra with single-component models, 
either purely thermal or absorbed PL, as a zero-th order attempt. For all sources,
these simple fits were rejected with high statistical significance.

The XMM-{\it Newton} spectral data are shown  in Figs. \ref{fig4} 
and \ref{fig4bis} and compared with model spectra. The bottom panel in 
each figures displays the ratio of the observed to predicted flux as a function of
energy. No emission lines are evident in these spectra, if we exclude Mkn 231 and 
IRAS19254-7245. 
The only appreciable feature is a fairly sharp, well characterized peak at 
E$\sim 0.7-1$ keV, likely of thermal origin. We have modelled this component
with a thermal emission template using the routine MEKAL in XSPEC (Mewe et al. 1985),
whose free parameters are the temperature $kT$ and the normalization.
The plasma metallicity has been set, for simplicity, to the solar value. 
This has no effect in our estimate of the the plasma temperature,
the assumed metallicity essentially affecting only the normalization of the 
thermal spectrum.

In the rest of the present Section we discuss each source separately.
The best-fit parameters are summarized in Table 3.
More detailed spectral decompositions are discussed in Sect. \ref{discussion}.

\subsection{IRAS 19254-7245, {\sl the Superantennae}}  
  
From the analysis of the ASCA data (Imanishi \& Ueno 1999; Pappa, Georgantopoulos, 
\& Stewart 2000), two alternative fits were proposed for this object:  
{\it a)} unabsorbed power law model with flat photon index ($\Gamma\sim1$), 
and  {\it   b)} absorbed PL model with $\Gamma\simeq 1.8$ and 
N$_H\sim 10^{22}$ cm$^{-2}$.  No Fe-K lines were detected by ASCA.  

XMM-Newton has detected a strong X-ray flux from IRAS 19254-7245. As shown in Fig. 
\ref{fig1}, the emission is centered in the southern nucleus of the interacting
pair, while no hard X-ray emission is apparently related to the northern one.
This is consistent with the results from observations at other wavelengths, 
which show that the southern nucleus, classified as a Seyfert 2, is always brighter
than the nothern one (Melnick \& Mirabel 1990).   By taking into account the EPIC PSF, 
the X-ray source appears to be significantly extended at low-energies (E$<2$ keV).
A detailed analysis of this complex source, resembling in many aspects the
prototypical type-II quasar NGC 6240, is reported in Braito et al. (2003a).

A single unabsorbed power-law model is not a good description of the XMM-{\it Newton} 
data because it seriously misfits the spectrum at low energies.  
By adding a thermal component to their composite model, 
Braito et al. (2003a) have obtained a temperature
kT=$0.85^{+0.1}_{-0.1}$ keV for the thermal component and $\Gamma=1.84\pm 0.20$
(N$_H=4.7\times10^{21}$cm$^{-2}$) for the hard PL component.  
The soft X-ray luminosity is $\sim 1.8\times 10^{42}$ erg s$^{-1}$.  
Braito et al. (2003a) have also found evidence for a Fe-K line at 6.4 keV, whose 
large equivalent width (EW$\simeq 1.4$ keV), together with the flat photon index, indicates 
that this object is a ``Compton thick" AGN, in which the detected hard X-ray emission 
is due to a pure reflected component plus a scattered one. Hence, the intrinsic AGN luminosity is
much higher than the observed value of $\sim 10^{42}$ erg s$^{-1}$, and
likely $\geq 10^{44}$ erg s$^{-1}$.

\begin{table*}  
\begin{center}  
\leavevmode  
\footnotesize  
\caption{Source Fluxes and Luminosities.
}  
\label{tab2}
\begin{tabular}[h]{lccccc}  
\hline \\[-5pt]  
    Name                 & CTS$^{(a)}$ &$S_{0.5-2}^{(b)}$          &$S_{2-10}^{(b)}$ & $L^{(c)}_{0.5-2}$ & $L^{(c)}_{2-10}$ \\[+5pt]  
                   &PN+MOS1+MOS2 &[$erg\ cm^{-2} s^{-1}]$& [$erg\ cm^{-2} s^{-1}]$& [$erg\ s^{-1}]$& $[erg\ s^{-1}$] \\   
\hline 
IRAS 12112$+$0305  &256$\pm 21$ &$1.2\times 10^{-14}$    & $1.5 \times10^{-14}$ & $3.4\times 10^{41}$ & $3.6\times 10^{41}$\\  
MKN231             & $2535\pm 51$ & $1.1\times 10^{-13}$ & $6.5 \times10^{-13}$ &$ 2.2\times 10^{42}$ & $6.4\times 10^{42}$ \\  
IRAS 14348$-$1447  &$297\pm 24$ &$1.8\times 10^{-14}$    & $2.0\times 10^{-14}$ &$ 8.9\times 10^{41}$ & $5.5 \times10^{41}$ \\  
IRAS 15250$+$3609  &$433\pm 26$ &$2.1\times 10^{-14}$  & $2.4\times 10^{-14}$ &$ 3.0\times 10^{41}$ & $3.0\times 10^{41}$ \\  
IRAS 17208$-$0014  & $401\pm 31$ &$2.3\times 10^{-14}$   & $4.0\times 10^{-14}$ &$ 6.0\times 10^{41}$ &$ 3.6\times 10^{41}$  \\  
IRAS 19254$-$7245  & $926\pm 33$&$5.3 \times10^{-14}$    & $2.3 \times10^{-13}$ &$ 1.8 \times10^{42}$ &$ 2.9\times 10^{42}$ \\  
IRAS 20100$-$4156  &$101\pm 17$ &$5.4\times 10^{-15}$    & $1.9\times 10^{-14}$ &$ 1.2\times 10^{42}$ &$ 1.6\times 10^{42}$  \\  
IRAS 20551$-$4250  & $942\pm 39$&$5.5\times10^{-14}$    & $1.4\times 10^{-13}$ &$ 4.2 \times10^{42}$ &$ 7.9\times 10^{42}$  \\   
IRAS 22491$-$1808  &$232\pm 19$ &$1.0\times10^{-14}$    & $6.4\times 10^{-15}$ &$ 1.6\times 10^{41}$ &$ 1.7\times 10^{41}$  \\  
IRAS 23128$-$5919  &$833\pm 34$ &$4.95\times10^{-14}$    & $1.7\times 10^{-13}$ &$ 1.0\times 10^{42}$ &$ 1.8\times 10^{42}$  \\  
\hline \\  
\end{tabular} 
\end{center}  
  $^{(a)}$ total net counts used in the spectral fits.  \\
  $^{(b)}$ observed fluxes.  \\
  $^{(c)}$ luminosities are corrected for galactic and intrinsic (N$_{\mathrm{H}}$)
		absorption.

\end{table*}

\subsection{MKN 231}  
  
Among local ULIRGs, Mkn231 is the most luminous object in the IR (Soifer et al. 
1984) and one of the best studied at all wavelengths. 
Observed with many X-ray instruments (ROSAT, ASCA and \textit{Chandra}), 
its X-ray properties still remain puzzling.   
Evidences of a combined AGN and SB activity emerged from the
ROSAT and ASCA data (Imanishi and Ueno 1999; Turner 1999). 
However the flatness of the X-ray spectrum at energies above 2 keV and the lack of 
detection of any strong Fe lines (Maloney and Reynolds 2000) appeared unusual.
These results have been confirmed with a recent Chandra observation
($\Gamma=1.3$, FeK line with EW$<$188 eV; see Gallagher et al. 2002). 
 
Our XMM-{\it Newton} field containing the source is shown in Fig.~\ref{fig2}:
the source appears very bright and extended. This extension is significantly
in excess of the PSF only for low-energy photons (E$<2$ keV), with a total
diameter of $\sim 1$ arcminute ($\sim 50$ kpc), while for higher energy photons
the source spatial profile is consistent with the PSF.
A detailed analysis of the XMM-{\it Newton} data on Mkn 231, plus those from a deep 
Beppo-SAX observation extending up to 80 keV, is reported in a companion
paper by Braito et al. (2003b).
Our results also confirm the very flat spectrum and the
detection of a moderately intense (EW$\sim$200 eV) Fe-K$\alpha$ line 
at 6.4 keV. The luminosity of the hard component is $\sim 3\times10^{42}$ erg s$^{-1}$.  
The presence of a highly obscured AGN is then confirmed by the BeppoSAX 
PDS data, but our XMM-{\it Newton} observations suggest the concomitant presence
of an important starburst component.

\subsection{IRAS 20551-4250 and IRAS 23128-5919}\label{I20551}
  
IRAS 20551-4250 has been previously detected with ASCA (Misaki et al. 1999), but at a  
very low significance level. XMM-{\it Newton} detects it with a high enough S/N to allow 
us to perform a detailed spectral analysis. The source appears as unresolved
by EPIC (see Figs.~\ref{fig2}).

A good fit (Fig.~\ref{fig4bis}) is obtained only with a three components model: 
a thermal model (kT=0.68$^{+0.12}_{-0.07}$ keV), and a ``leaky-absorber" 
continuum, including an absorbed plus a non-absorbed power law 
spectrum with the same photon index. We found that a good fit can be obtained 
with $\Gamma=1.8^{+0.2}_{-0.2}$ and N$_H=8 \times10^{23}$cm$^{-2}$.  
The 6.4 keV Fe-K line is not detected, but the upper limit on its equivalent 
width ($\leq 1$ keV) is consistent with the N$_H$ value obtained from the fit.  
The intrinsic 2-10 keV luminosity of this object, corrected for absorption, 
is $\sim 7.0\times 10^{42}$ 
erg s$^{-1}$, while the soft X-ray luminosity is $\sim 4.2\times 10^{42}$ erg s$^{-1}$. 
The high X-ray luminosity and the properties of the X-ray spectrum clearly 
suggest the presence of an extinguished AGN.

The XMM-{\it Newton} observations of IRAS 23128-5919 have been strongly affected by 
high background (only 9.4 ksec of good data are available), and this prevented us
from attempting detailed spectral fits. The source appears as point-like and
relatively bright. The spectrum is reasonably well reproduced by the sum of
a thermal model and a simple power-law with no absorption. However, the power-law's
spectral index in this solution is unphysically hard ($\Gamma=1.04\pm 0.13$), 
suggesting the presence of an absorbed component. We then adopted
the same ``leaky-absorber" model as discussed for IRAS 20551-4250, and found best-fit
parameters kT=0.65$\pm 0.15$ keV, $\Gamma=1.67^{+0.09}_{-0.34}$ and
N$_H=6.9\pm 3.5 \times10^{22}$cm$^{-2}$. The luminosities of the thermal
and de-absorbed power-law components are $\sim 1.5\times 10^{41}$ and 
$\sim 2.7\times 10^{42}$erg s$^{-1}$. The large gas column density and power-law 
luminosity clearly indicate the presence of an AGN, whereas a spectroscopic IR
study by Charmandaris et al. (2002) revealed the predominance of a starburst.
Our subsequent analysis in Sect. \ref{spectral} suggests that IRAS 23128-5919 is 
a transition object, with properties intermediate between those of an AGN and a SB.

\subsection{IRAS 14348-1447, IRAS 15250+3609, and IRAS 17208-0014}  

All these sources have intermediate-quality detections.
While the latter two are consistent with a point source emission at all energies,
IRAS 14348-1447 appears to be resolved in a bow-like structure of total size
$\sim$30 arcsec in the North-South direction. The close-up image in
Fig. \ref{fig3} shows also the presence of another relatively bright
($S_{2-10keV}\sim 5\times 10^{-15}$ [erg/cm$^2$/s]) source located only 20 arcsec 
away in the SE direction, a position in which no optical counterpart is detectable
in the POSS image.

IRAS 14348-1447 and IRAS 17208-0014 show evidence of a single-temperature plasma 
emission with kT$\simeq 0.62{\pm 0.08}$ keV and kT$\simeq 0.75{\pm 0.08}$ keV respectively. 
The spectrum of IRAS 15250+3609, instead, is unique among those from our
spectral survey in requiring the presence of multi-temperature plasmas
(a two-temperature solution indicates kT$\simeq 0.26{\pm 0.09}$ keV and
kT$\simeq 0.64{\pm 0.15}$ keV).

In addition to these thermal emissions we find a very significant hard X-ray component 
in all three
sources. Acceptable spectral fits have then been obtained by combining a Mekal-modeled 
thermal emissions
and an absorbed PL having $\Gamma \sim 2.2$, except for IRAS15250+3609 whose 
photon-index is flatter ($\Gamma\sim 1.2$) (see Table \ref{tab3}). 
In all sources some absorption in excess of the galactic value
is required ($N_{\mathrm{H}}> 10^{21}$cm$^{-2}$). All three sources have X-ray 
luminosities in the range $10^{41}- 10^{42}$ erg s$^{-1}$.

\begin{table*}  
\begin{center}  
\leavevmode  
\caption{Best-Fit Spectral Parameters. }
\label{tab22}
\begin{tabular}[h]{lccccccc}  
\hline \\[-5pt]  
    Name           & MODEL$^{(c)}$& $\chi^2/\nu$&kT 		  & L$^{(d)}_{thermal}$ 	& $\Gamma$ & N$_H$    & L$^{(d)}_{PL}$\\[+5pt]  
  		   &      &		 & 	  & 				&      &           &L$^{(d)}_{binaries}$ \\[+5pt]  
                   & 	  & &[keV] 	  &[$erg\ s^{-1}$]	&          &$[cm^{-2}]$& [$erg\ s^{-1}$]  \\   
\hline 
IRAS 12112$+$0305  &A	&$14.9/17$ & 0.79$^{+0.43}_{-0.25}$      & $0.8\times 10^{41}$  & 1.92$^{+0.33}_{-0.55}$ & 0.8$^{+1.4}_ {-0.8}\times 10^{21}$& $6.3\times 10^{41}$ \\ 
                   &B	&$16.3/18$ & 0.80$^{+0.24}_{-0.16}$      & $1.1 \times10^{41}$  & 1.1$^\star$& / & $6.1 \times 10^{41}$ \\ 

  		   &      &		 & 	  & 				&      &           &   \\ 

MKN231$^{(b)}$     & D$^{(f)}$    &  196.8/156 & $0.96^{+0.09}_{-0.14}$   & $3.7 \times10^{41}$           &  $1.30^{+0.16}_{-0.15}$    & $7.39^{+0.19}_ {-0.91} \times 10^{21}$           &   $8.3 \times 10^{42}$                      \\  
  		   &      &		 & $0.37^{+0.09}_{-0.05} $	  & 				&      &           &   \\ 
  		   &      &		 & 	  & 				&      &           &   \\ 

IRAS 14348$-$1447  & A & 10.3/15&0.61$^{+0.20}_{-0.19}$      & $3.5\times 10^{41}$  & 2.18$^{+0.51}_{-0.60}$ & 0.3$^{+0.7}_{-0.3} \times10^{22}$ & $1.3\times 10^{42}$ \\   
		   & B & 13.7/16&0.64$^{+0.16}_{-0.13} $     & $3.7 \times10^{41}$  & 1.1$^\star$ 		   & 0.5  $^{+2.6}_{-0.5} \times 10^{21}$& $1.1\times 10^{42}$ \\ 

  		   &      &		 & 	  & 				&      &           &   \\ 

IRAS 15250$+$3609$^{(a)}$& A& $13.9/18$&0.64$^{+0.14}_{-0.20}$& $2.3 \times10^{41}$  & 1.21$^{+0.62}_{-0.66}$ &     /       & $3.9 \times10^{41}$\\  
                               & & &0.26$^{+0.08}_ {-0.07}$                     &      &             &                            \\
                   &B&  $14.9/18$ &0.66$^{+0.17}_{-0.10}$& $2.2  \times10^{41}$  & 1.1$^\star$ &     /       &  $3.2 \times10^{41}$\\  
                               & & &0.26$^{+0.07}_ {-0.07}$                     &      &             &                             \\

  		   &      &		 & 	  & 				&      &           &   \\ 

IRAS 17208$-$0014  & A&$16.3/24$&0.76 $^{+0.11}_{-0.12}$     & $1.6  \times10^{41}$  & 2.26$^{+1.19}_{-0.98} $ & 1.1 $^{+1.4}_{-1.1} \times10^{22}$ & $8.1\times  10^{41}$\\  
                  & B&$17.1/24$&0.74 $^{+0.14}_{-0.11}$     & $1.3  \times10^{41}$  & 1.30$^{+0.10}_{-0.60} $ & 2.6 $^{ +7.2}_{-2.6} \times10^{21}$& $5.3 \times10^{41}$\\  
  		   &      &		 & 	  & 				&      &           &   \\ 

IRAS 19254$-$7245$^{(b)}$  & C& $48.1/56$& 0.85$^{+0.14}_{-0.09}$&$4.5\times 10^{41}$ &1.84$^{+0.60}_{-0.55}$&0.47$^{ +0.52}_{-0.25} \times10^{22}$ &$5.2 \times10^{42}$\\  

  		   &      &		 & 	  & 				&      &           &   \\ 

IRAS 20100$-$4156  & A &  $4.8/8$ &0.75$^{+0.24}_{-0.32}$  & $4.1  \times10^{41}$  & 1.7$^\star$ 	 & 2.6$^{+ 4.9}_{-2.6}\times10^{22}$ & $2.3 \times10^{42}$   \\  
		   & B &  $4.7/8$ &0.75$^{+0.23}_{-0.26}$  & $4.1  \times10^{41}$  & 1.1$^\star$ & 2.2$^{+4.8}_{-2.2} \times10^{22}$  & $2.0\times 10^{42}$ \\  
  		   &      &		 & 	  & 				&      &           &   \\ 

IRAS 20551$-$4250   & E$^{(f)}$&$45.7/49$& 0.66$^{+0.10}_{-0.06}$   & $3.0\times 10^{41}$  & 1.8$^{+0.20}_{-0.19}$  & 7.9$^{+6.9}_{-1.9}\times 10^{23}$&  $7.0 \times10^{42}$  \\   
  		   &      &		 & 	  & 				&      &           &   \\ 
IRAS 22491$-$1808  & A& $22.9/21$  &0.69$^{+0.19}_{-0.21}$ & $1.5 \times10^{41}$  & 1.98$^{+0.67}_{-0.63}$ &     /    &  $3.6 \times10^{41}$  \\  
                   & B& $26.7/23$  &0.66$^{+0.15}_{-0.20}$ & $1.8 \times10^{41}$  & 1.1$^\star$ &     /       &  $3.8  \times10^{41}$   \\  
  		   &      &		 & 	  & 				&      &           &   \\ 

IRAS 23128$-$5919  & E$^{(f)}$& $32.7/44$  &0.65$^{+0.15}_{-0.14}$ & $1.5\times 10^{41}$  & 1.67$^{+0.09}_{-0.34}$    & 6.9$^{+4.7}_{-3.4}\times 10^{22}$ & $2.7 \times10^{42}$ \\  
\hline \\  

\end{tabular} 
\end{center}  
\label{tab3}
$^{(a)}$: For this source an acceptable fit requires a two-temperature plasma.\\ 
$^{(b)}$: XMM{-\sl Newton} observations reported in Braito et al. (2003a and 2003b) \\
$^{(c)}$: Model A: Thermal emission plus an absorbed power-law model; Model B: thermal  
emission plus cutoff power-law; Model C: Thermal emission plus a Compton reflected and 
scattered countinum (see Braito et al. 2003a); Model 
D: two thermal emission plus a ``leaky-absorber''; Model 
E: thermal emission plus a ``leaky-absorber'' continuum.\\
$^{(d)}$: Luminosities are between 0.5 and 10 keV.\\
$^{(f)}$: The covering factor for the  ``leaky-absorber'' continuum are 9\%, 5\%
and 54\% for MKN231, IRAS 20551$-$4250 and IRAS 23128$-$5919 respectively.\\
N.B. The symbol  $^\star$ indicate that  the parameter has been kept fixed.

\end{table*}

\begin{figure*}
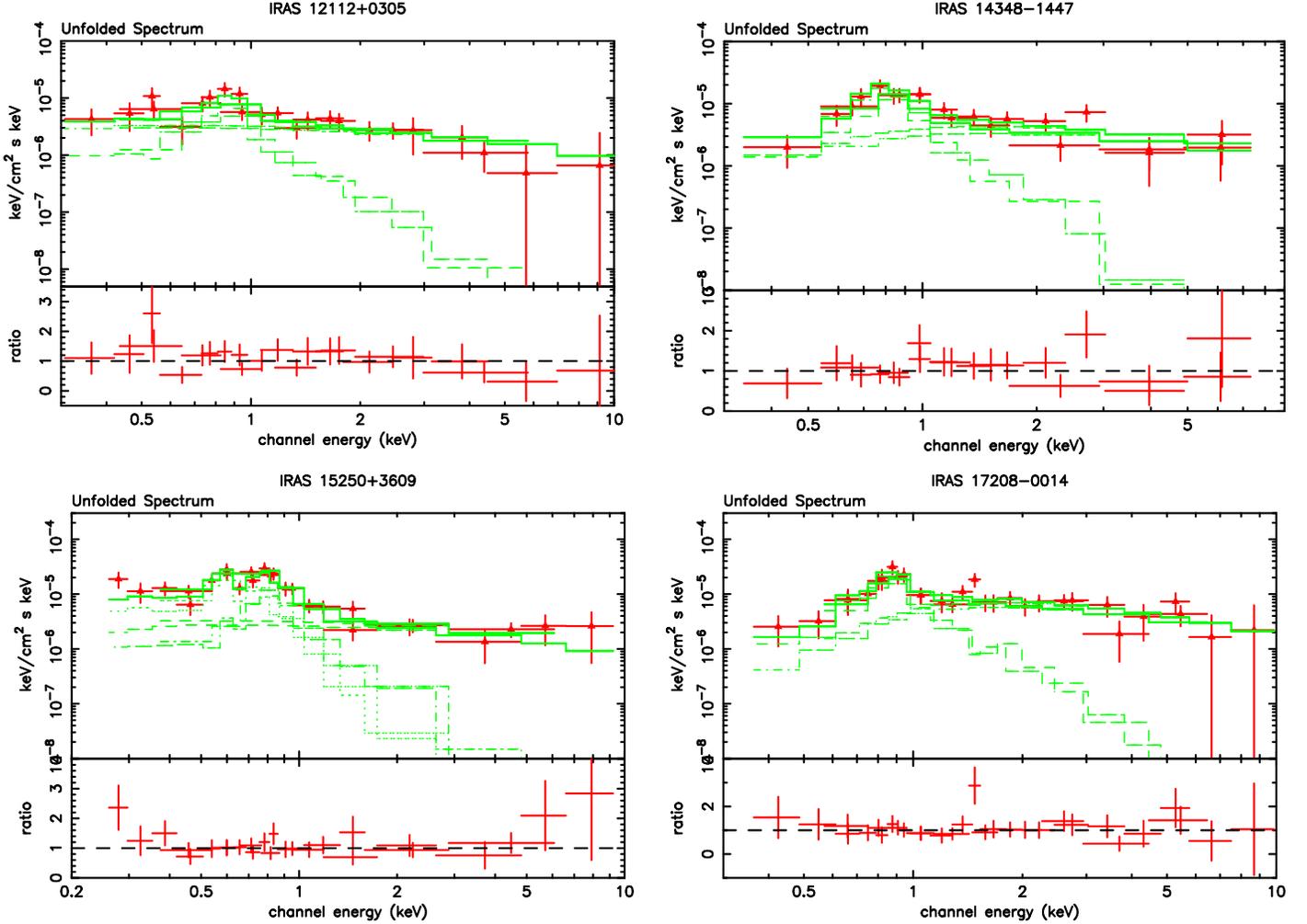

\begin{tabular}{cc}
\hskip-0.5truecm\epsfig{file=IRAS12112_bin.ps,angle=-90,width=9cm           }&
\hskip-0.truecm\epsfig{file=IRAS14348_bin.ps,angle=-90,width=9cm           }\\
\hskip-0.2truecm\epsfig{file=IRAS15250_bin.ps,angle=-90,width=9cm           }&
\hskip-0.truecm\epsfig{file=IRAS17208_bin.ps,angle=-90,width=9cm           }\\
\end{tabular}
\caption{XMM-{\it Newton} spectra in energy units and ratios of data to the best-fit model
values as a function of energy, for the sources IRAS12112+0305, IRAS14348-1447, 
IRAS15250+3609, and IRAS17208-0014, from top left to bottom right.}
\label{fig4}
\end{figure*}

\begin{figure*}
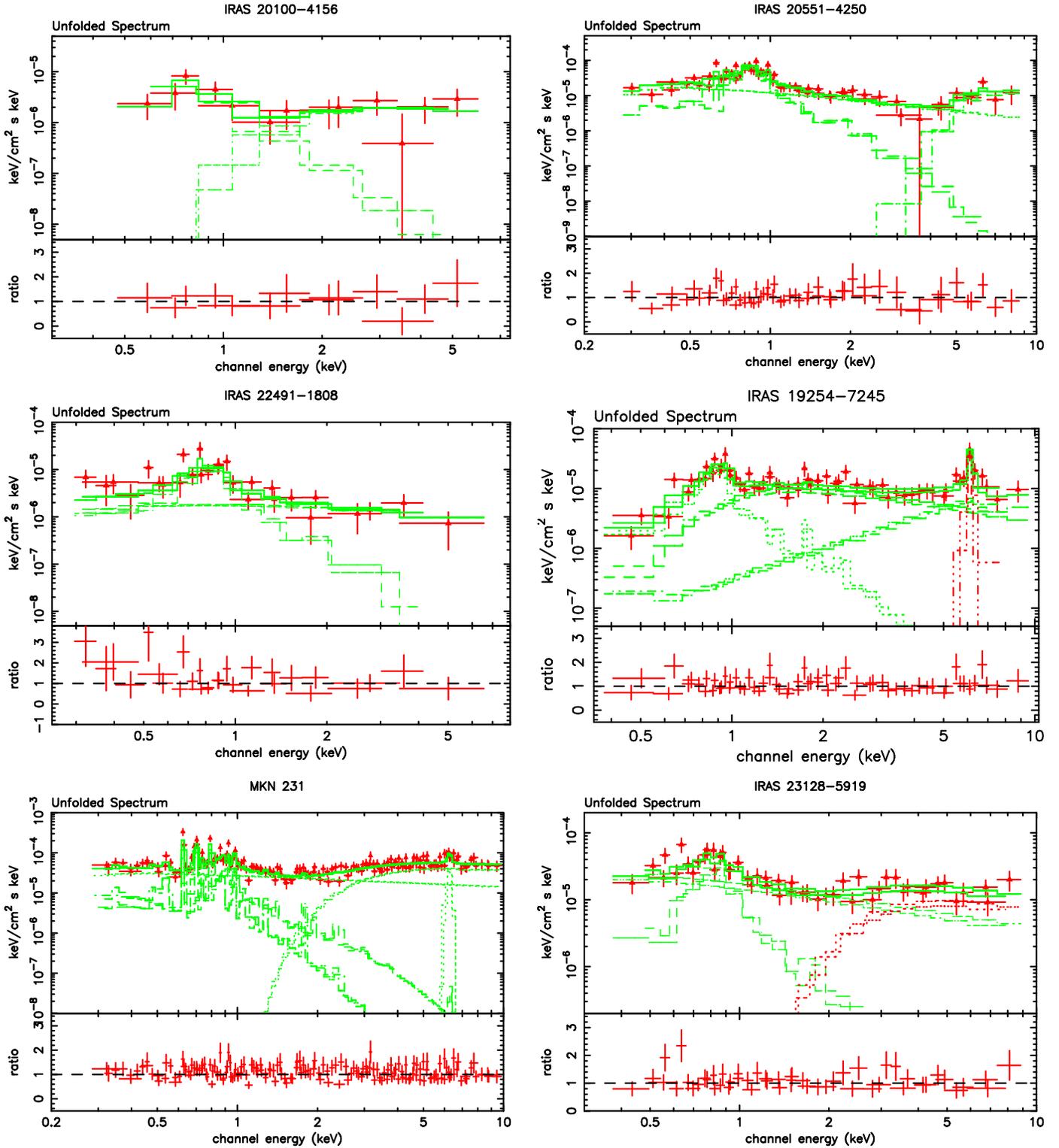

\begin{tabular}{cc}
\hskip-0.5truecm\epsfig{file=IRAS20100_bin.ps,angle=-90,width=9cm   }&
\hskip-0.truecm\epsfig{file=IRAS20551_spec.ps,angle=-90,width=9cm   }\\
\hskip-0.5truecm\epsfig{file=IRAS22491_bin.ps,angle=-90,width=9cm   }&
\hskip-0.truecm\epsfig{file=IRAS19254_spec.ps,angle=-90,width=9cm   }\\
\hskip-0.5truecm\epsfig{file=mkn231_spec.ps,angle=-90,width=9cm     }&
\hskip-0.truecm\epsfig{file=iras_23128_spec.ps,angle=-90,width=9cm  } \\
\end{tabular}
\caption{XMM-{\it Newton} spectra in physical units and ratios of data to the best-fit model,
for the sources IRAS20100-4156, IRAS 20551-4250, IRAS 22491-1808, IRAS 19254-7245,
Mkn 231 and IRAS 23128-5919.}
\label{fig4bis}
\end{figure*}

\subsection{IRAS 12112+0305, IRAS 20100-4156 and IRAS 22491-1808 }  
  
All these sources are detected at the lowest level of significance. 
This has prevented us from performing any detailed spectral analyses. 
We only tried on these sources two separate fits with a thermal model and 
an absorbed PL. For all three sources, a single thermal or a PL
model are rejected by the data. 
Successful spectral fits require a combination of thermal and PL emissions
(see Table \ref{tab3}).

The intrinsic hard ($2-10$ keV) X-ray luminosities are  $1.7\times  10^{41}$erg s$^{-1}$ 
for IRAS 22491-1808, $\sim 1.6 \times 10^{42}$erg s$^{-1}$ for IRAS 20100-4156,
and $3.6\times \ 10^{41}$erg s$^{-1}$ for IRAS 12112+0305.

\section{THE SPECTRAL ANALYSIS}\label{discussion}

The X-ray spectral properties of the ULIRG sample appear to be fairly uniform. 
The spectra are well reproduced by the combination of a soft thermal 
and a hard PL components. For a physical interpretation of the ULIRG phenomenon,
the implications of these results are briefly
discussed in this and the next Section. A more detailed comparison of these
XMM-{\it Newton} spectra with model predictions is deferred to Persic et al. (2003a).

\subsection{Evidence for AGN-dominated emission}

One of the motivations for the present survey has been the opportunity offered by hard
X-ray data to constrain the relative AGN and starburst contributions to the source 
activity. A dominant AGN contribution should be detectable in X-rays as either 
{\sl a)} an high-luminosity X-ray emission, L$_{2-10 keV}>10^{42}\ erg/s$; 
or {\sl b)} highly extinguished hard X-ray components with N$_H>10^{22}\ 
cm^{-2}$, as revealed by very flat or inverted hard X-ray spectra; 
or {\sl c)} Fe-K complexes at $\sim$6.4 keV with large Equivalent Widths 
($EW\sim$1 keV), corresponding to Iron fluorescent emission by cold molecular
material   illuminated by the AGN's energetic radiation
field.

We found such evidence for an AGN spectrum in 4 out of the 10 sources observed.
The most obvious AGN is IRAS 19254-7245, which has 
both a strong Fe-K line and a very flat (probably scattered) hard continuum, 
implying an intrinsic luminosity $L_{2-10}>10^{44}\ erg/s$ (Braito et al. 2003a).
 
The ULIRG Mkn 231 has been clearly identified from optical/IR observations as 
an AGN-dominated source, as inferred in particular from the flat, almost power-law,
mid-IR continuum observed by Genzel et al. (1998) and the low EW of the PAH emissions.
The source is also classified as a Broad Absorption Line quasar from optical
spectroscopy (Smith et al.1995).
The XMM-{\it Newton} observations reveal a moderate luminosity X-ray source with a very
flat spectrum ($\Gamma\simeq 0.9$) between 1 and 10 keV, while Beppo-SAX data reported
by Braito et al. (2003b) reveal an highly extinguished (N$_H>10^{24} cm^{-2}$)
component above 10 keV. These data fully confirm the classification of Mkn 231
as a dusty and extinguished AGN.

IRAS 20551-4250 is a third ULIRG with evidence from the XMM-{\it Newton} spectrum of harboring 
a high-column-density (N$_H\sim 10^{24}cm^{-2}$) extinguished AGN, and a similar
evidence for the presence of a low-luminosity AGN was found in IRAS 23128-5919.
Neither these two sources nor IRAS 19254-7245 showed evidence for AGN 
activity from infrared spectroscopy (Genzel et al. 1998), which illustrates the
useful complementarity between the IR and the X-ray approach. A detailed 
spectrophotometric
study of Superantennae by Berta et al. (2003) has found evidence, however, for a 
combined AGN and starburst emissions with comparable IR bolometric luminosities.

Further indications about the relative AGN/starburst contributions will come from the
comparison of soft X-ray thermal and hard X-ray power-law emissions discussed in
Sect. \ref{spectral}.

\subsection{Starburst emission}\label{SB}

The spectra of local SB galaxies in the 0.5-10 keV band can be described 
as a combination of warm thermal emission (with typically $kT \simeq 0.6-0.8$ keV) 
dominating at energies $\leq 1$ keV, plus a harder spectrum producing the bulk of
the 2-10 keV flux.  

The warm thermal component is interpreted as originating from the boundary region  
between the hot, tenuous outgoing galactic wind and the cool, dense ISM (e.g.,  
Strickland et al. 2000).  

\begin{figure*}
\begin{tabular}{cc}
\hskip-0.5truecm\epsfig{file=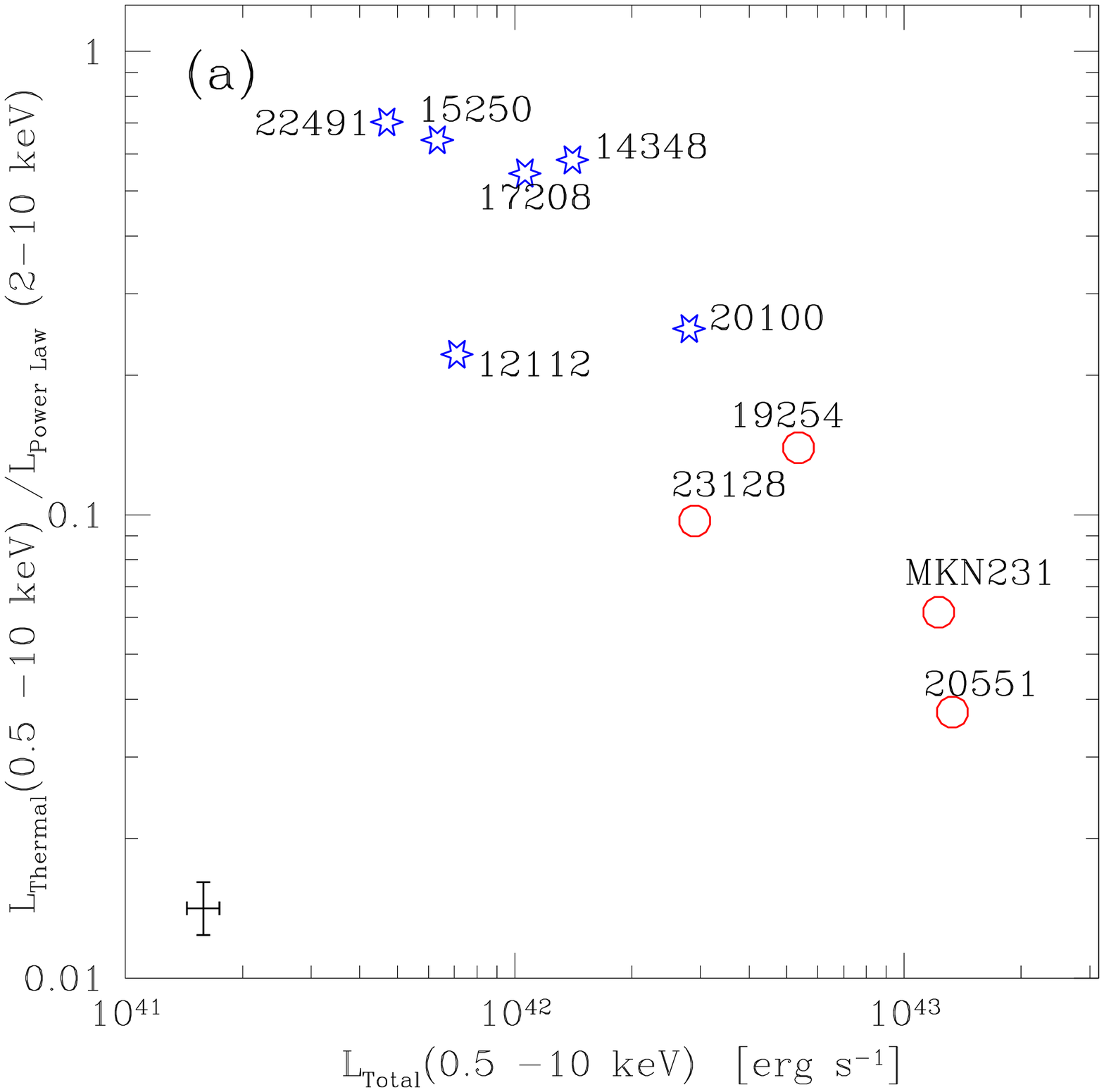,width=0.40\textwidth}&
\hskip-0.2truecm\epsfig{file=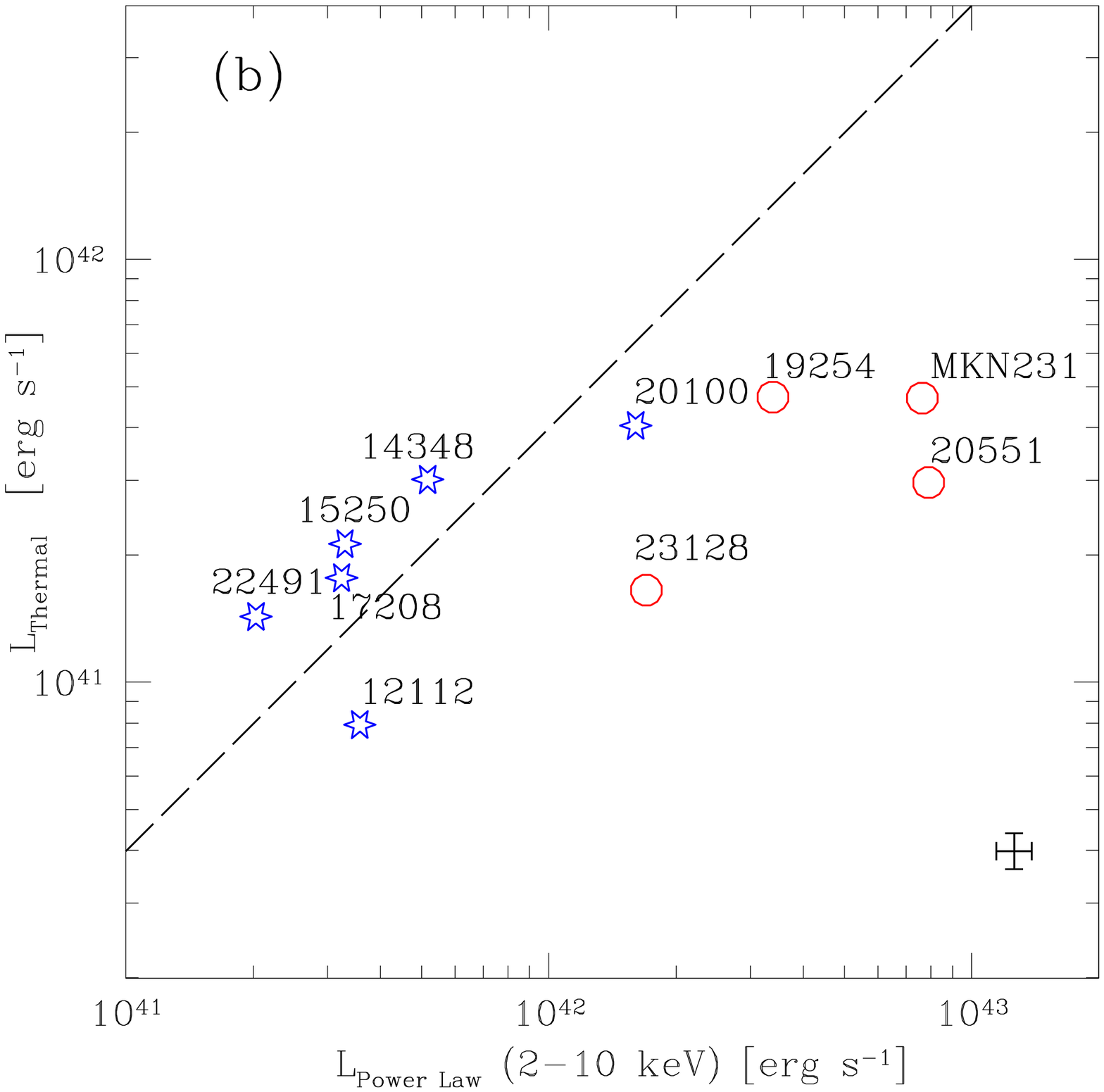,width=0.40\textwidth}\\
\hskip-0.2truecm\epsfig{file=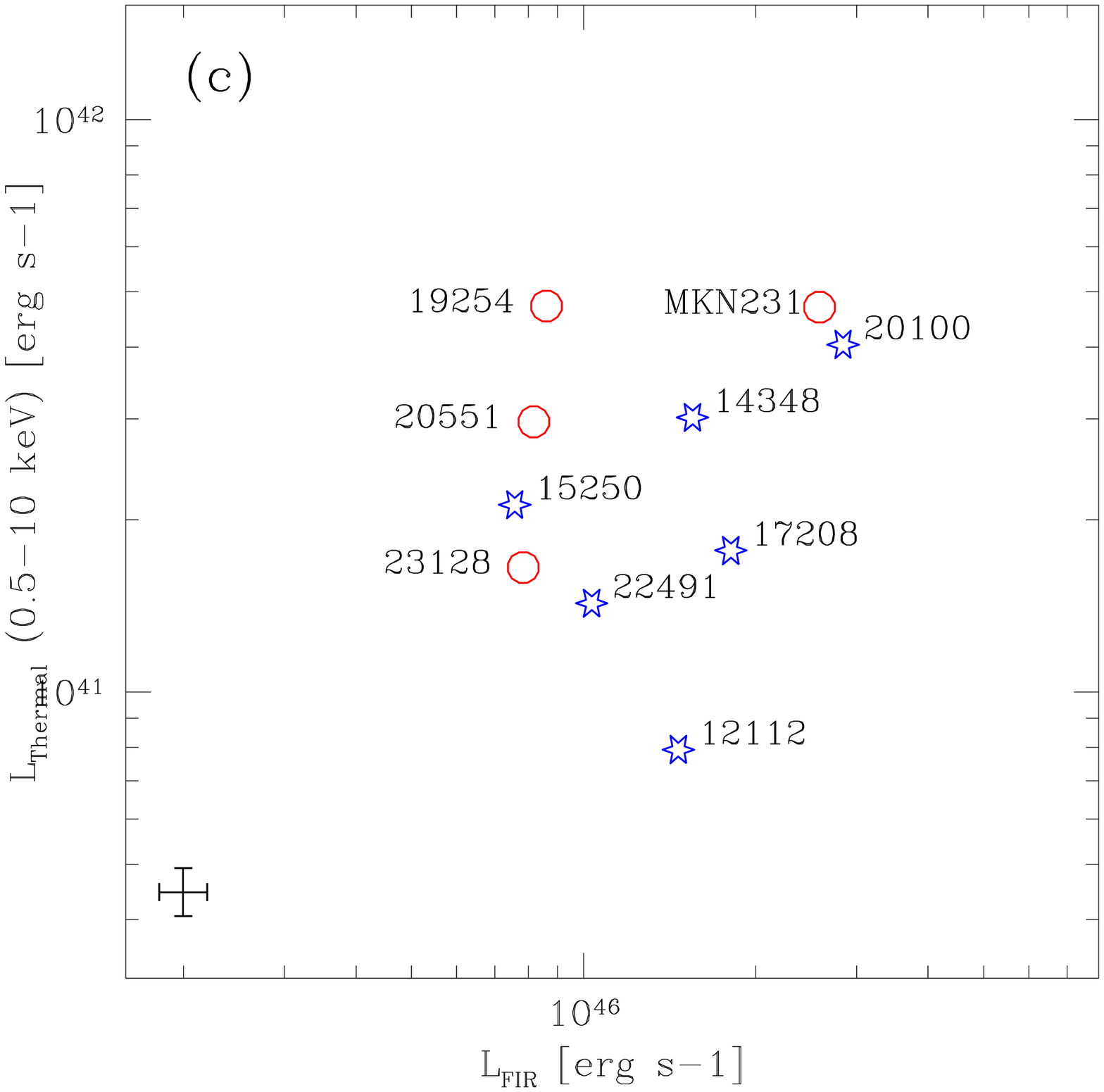,      width=0.40\textwidth}&
\hskip-0.5truecm\epsfig{file=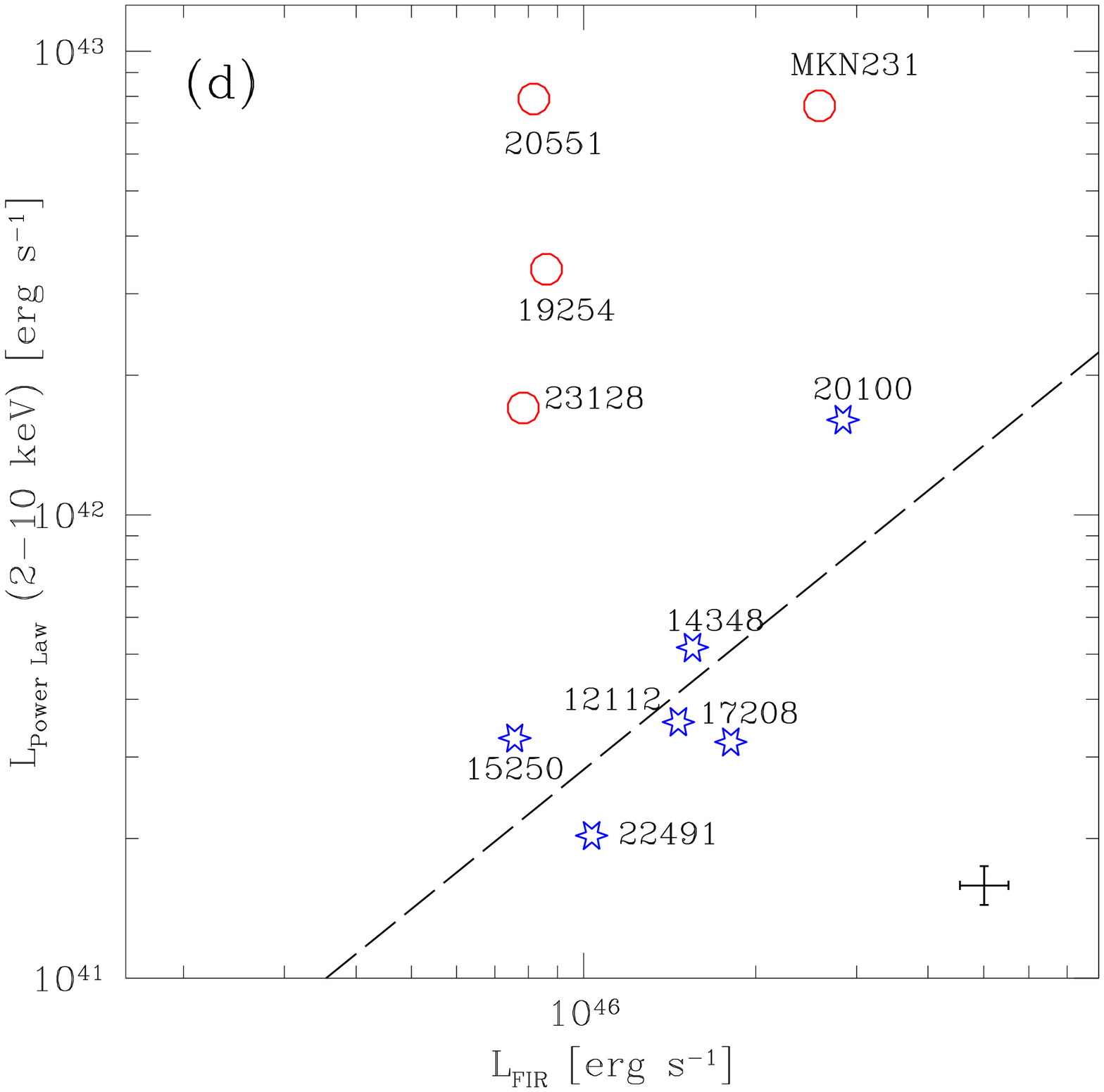,width=0.40\textwidth}\\
\hskip-0.5truecm\epsfig{file=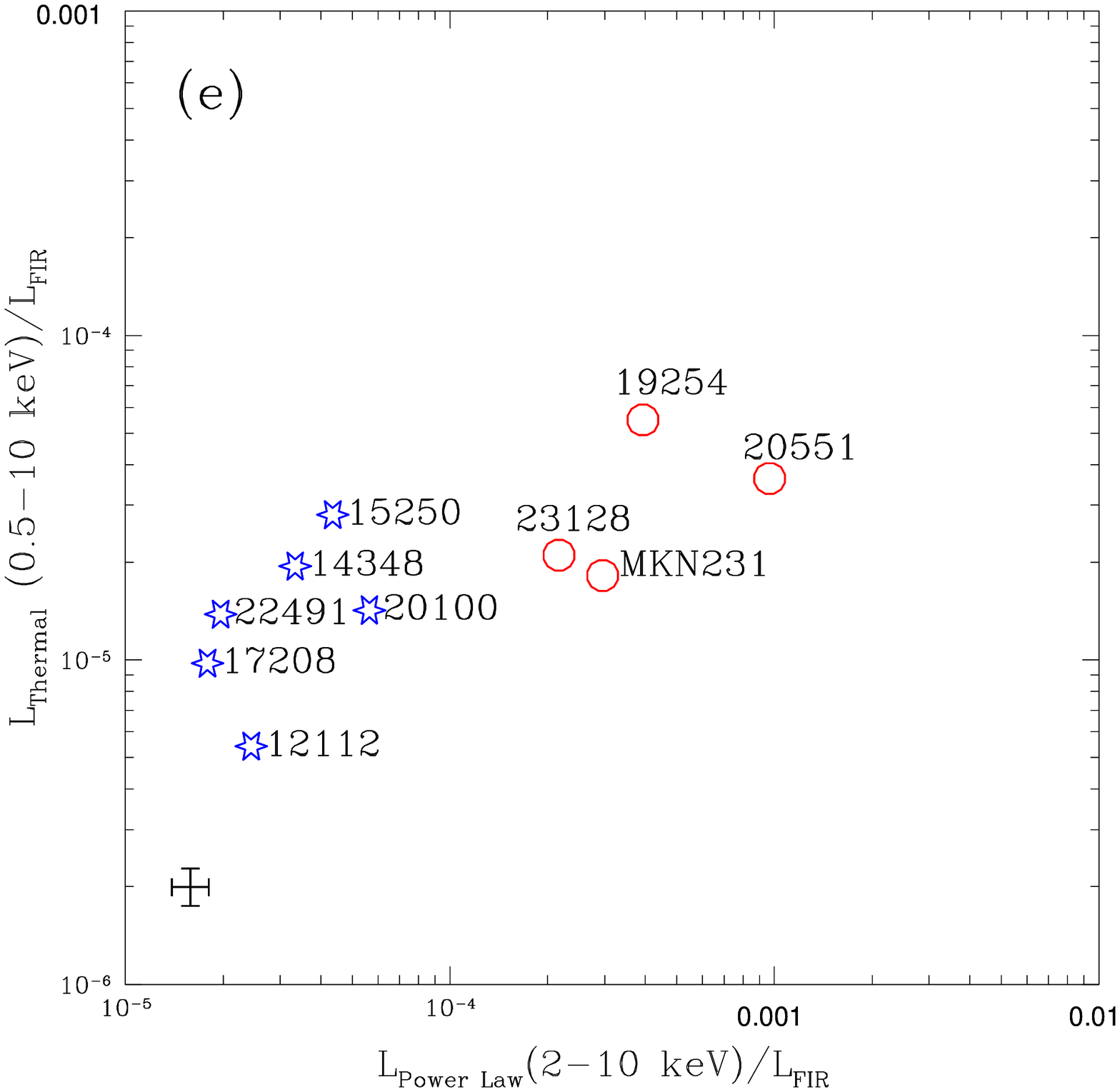,width=0.40\textwidth}
\\
\end{tabular}
\caption{Luminosity-luminosity plots of various emission components identified in the 
XMM-{\it Newton} spectra of the ULIRG sample. Starred symbols refer to sources that we 
classify as dominated by starbursts, open circles to AGN-dominated sources. 
Panel {\sl (a)}: luminosity ratio between the thermal and the power-law 
(2-10 keV) component after correction for intrinsic absorption, versus the bolometric 
X-ray luminosity measured in the total 0.5-10 keV band. 
Panel {\sl (b)}: luminosity-luminosity plot of the thermal versus PL components.
Panels {\sl (c)} and {\sl (d)}: luminosities of the thermal and PL components 
against the bolometric far-IR luminosities L$_{FIR}$.
Panel {\sl (e)}: the ratios of the thermal and PL luminosities to the bolometric
far-IR luminosity L$_{FIR}$.
Dashed lines indicate proportionality relations.
}
\label{fig_comp}
\end{figure*}

The harder component has various interpretations in the literature 
       \footnote{Various authors have analysed X-ray data on specific nearby 
	   starbursts: Ptak et al. (1997) and Cappi et al. (1999) for NGC 253 and M 82; 
	   Okada et al. (1997) for M83; Della Ceca et al. (1996) for NGC 1569;  
	   Della Ceca et al. (1999) for NGC 2146; Moran et al. (1999) for NGC 3256;  Della Ceca et al 1997 for NGC449;
	   Zezas et al. (1998) and Della Ceca et al. (2002) for NGC 3310, NGC 3690 (Arp 299). 
	   See also Dahlem et al. (1998) for a review.}, 
either in terms of a very hot (kT$\geq$5 keV) thermal or a $\Gamma \sim 2$ PL model,
but no definitive conclusions were reached, as the thermal and PL fits are, in 
general, similarly successful. 

Recently, Persic \& Rephaeli (2002; see also David, Jones \& Forman 1992) 
have developed a detailed quantitative model of synthetic X-ray spectra of 
SB galaxies, based on evolutionary populations of galactic stars and 
adopting template X-ray spectra for the relevant emission processes. 
They suggested that high- and low-mass X-ray binaries (HMXBs, LMXBs) 
contribute most of the 2-15 keV spectrum, in the absence of AGN emission.  
Both categories of X-ray binaries have spectra that can be described as variously cutoff 
PLs (White et al. 1983; Christian \& Swank 1997). For example, in the case of one  
isolated episode of intense star formation, the 2-10 keV emission of the starburst  
would be dominated by HMXBs, and would be described as a $\Gamma=1.2$ power  
law.      In more moderate starbursts (such as those observed in local SB galaxies), 
in addition to the HMXBs related to the SB proper, also the LMXB population in the 
underlying disk is expected to provide important contributions
to the X-ray emission. In the case of a mix of HMXBs and LMXBs of various luminosities  
with Galactic proportions, the 2-10 keV emission can be described as a cutoff PL of the 
form $E^{-\Gamma} exp^{-E/kT}$, with photon index $\Gamma \sim 1.1$ and cutoff  
energy $kT \sim 8$ keV (Persic \& Rephaeli 2002). This spectrum is to be corrected 
for the intrinsic absorption with variable column density N$_H$.

\begin{table} 
  \caption{Observed ULIRGs Classification}  
  \label{tab3}  
  \begin{center}  
    \leavevmode  
    \footnotesize  
\begin{tabular}{lccc}  
\hline  
\bf{Name }        &\bf{Optical} & \bf{Mid-IR}& \bf{X-ray} \\      
 (1) &(2)     &(3)                     &(4)       \\   
\hline  
IRAS 12112+0305 &    L     & SB       & SB\\  
Mkn 231         &   AGN-1  & AGN      & AGN   \\   
IRAS 14348-1447 &    L     & SB       &  SB\\  
IRAS 15250+3609 &    L     & SB       &  SB\\  
IRAS 17208-0014 &    HII   & SB       &  SB\\  
IRAS 19254-7245 &  AGN-2   & SB & AGN   \\  
IRAS 20100-4156 &   HII    & SB & SB/AGN\\  
IRAS 20551-4250 &   HII    & SB & AGN   \\   
IRAS 22491-1808 &   HII    & SB & SB    \\  
IRAS 23128-5919 &   HII    & SB & SB/AGN  \\  
\hline   
\end{tabular}  
\end{center}  
\footnotesize\textit{{Note. $-$ 
Col.(1): Object name. 
Col.(2): Optical classification: Liner (L), AGN or HII region (Lutz et al. 1999; Veilleux et al. 
1999).  
Col.(3): Mid Infrared classification based on ISO  spectroscopy Starburst (SB) or AGN 
(Genzel et al. 1998). 
Col.(4): X-ray Classification.}}  
\end{table}

\subsection{Spectral decompositions}\label{spectral}

Following the above guidelines, we have tried to fit the spectra of our ULIRGs with 
various astrophysically motivated composites.
The first clear feature to consider is the ubiquitous low-energy excess emission
between 0.5 and 1 keV. One possibility to explain it would be
a reflected component by photoionized gas, as found for example in the
Seyfert 2 NGC 1068 (Kinkhabwala et al. 2002, Brinkman et al. 2002). 
We have tried to reproduce
the source spectra with such an emission, but in all cases have failed to obtain 
acceptable fits. In particular, for IRAS 19254-7245 the Fe-K line at 6.49 keV
detected by Braito et al. (2003a) implies the presence of an essentially cold 
neutral gas, which is unable to explain the low-energy emission features. For MKN 231, 
Braito et al. (2003b) attempt to explain it entirely with an ionized reflection, which
provides however a very poor overall fit, while a thermal ionized gas emission
is required by the data.
As for IRAS 20551-4250, an attempt to fit the low-energy spectrum with photo-ionized
emission is similarly unsuccessful, as it would require an unphysically
large value of the X-ray spectral index $\Gamma$ and would generate a quite poor 
overall fit.

Also in all other sample's sources the photo-ionized AGN emission model would
leave significant residuals in all cases, which implies that collisionally
ionized hot thermal gas is in any case required.
We have modelled this ubiquitous low-energy thermal component
with the Mekal/XSPEC code at constant (solar) metallicity
(for the source IRAS 15250+3609 a two-temperature component was required).

Then we have added a PL with photon index $\Gamma$, photoelectrically absorbed
through a column density N$_H$ to describe the hard X-ray excess that shows up at
energies $\geq 2$ keV. 
For IRAS 20551-4250 we have used the "leaky absorber" model (Sect. \ref{I20551}), 
while for Mkn 231 we have not considered in the present analysis the highly-extinguished 
(N$_H>10^{24}cm^{-2}$) component found in Beppo-SAX data by Braito et al. (2003b).
Alternatively, for all the sources we have attempted also to fit the high-energy data with 
an X-ray binary model with $\Gamma=1.1$, cutoff energy $kT \sim 8$ keV and variable
low-energy absorbing column density N$_H$ (see Sect. \ref{SB}).

The results of these spectral decompositions are reported in Table \ref{tab3} and shown in 
Figs. \ref{fig4} and \ref{fig4bis}. In Fig. \ref{fig_comp} 
we explore possible relationships between the best-fit parameters.
Fig.\ref{fig_comp}{\sl (a)} is a plot of the ratio between the luminosities of the 
thermal and PL components (after correction for intrinsic absorption), 
as a function of the  X-ray luminosity measured in the total 0.5-10 keV band. 
SB-dominated ULIRGs
(starred symbols) occupy a fairly well defined region with L$_{total}\sim 10^{42}erg/s$
and L$_{thermal}$/L$_{PL}\gsimeq 0.2$. IRAS 12112+0305 appears to have 
a very low value of L$_{thermal}$ compared with other starbursts.
The AGN-dominated sources have enhanced X-ray luminosities with 
L$_{thermal}$/L$_{PL}\leq 0.1$:
as expected, these AGN ULIRGs have an energetically-dominant power-law component.
In the following we use the symbol L$_{PL}$ to indicate the luminosity 
derived from the fit with a generic PL, rather than
the one corresponding to the best-fit X-ray binary model
(in any case the difference between the two is a few tens percent at most,
see Table \ref{tab3}).

Fig. \ref{fig_comp}{\sl (b)} compares the luminosities of the thermal and PL
components: a fairly clear correlation is apparent, with the
AGN population occupying the high-luminosity end in both quantities.  
Again the AGN-dominated sources show an excess of the PL component luminosity compared
with that of the thermal plasma.

Fig. \ref{fig_comp}{\sl (c)} plots the luminosities of the thermal component 
L$_{thermal}$ against the bolometric far-IR luminosities L$_{FIR}$, while 
Fig. \ref{fig_comp}{\sl (d)} does the same for the luminosities of the PL component. 
If we consider the starburst-dominated population, the L$_{PL}$ and 
L$_{FIR}$ luminosities appear both reasonably correlated, while the AGN-dominated
sources show similar L$_{FIR}$ values but largely enhanced X-ray PL emission.
On the contrary, we find a remarkable lack of correlation between L$_{FIR}$
and the emissivity of the thermal component.

Finally, panel \ref{fig_comp}{\sl (e)} shows that the SB-dominated
sources have fairly well defined L$_X$/L$_{FIR}$ ratios
for both the thermal and PL components. The AGN sources
display the usual excess of PL emission. On this regard, the sources IRAS
20100-4156 and 23128-5919 tend to occupy positions in the various plots which are intermediate 
between SB and AGN dominated populations, and may have genuinely
intermediate properties.

\begin{figure*}
\vskip -2.0truecm\epsfig{file=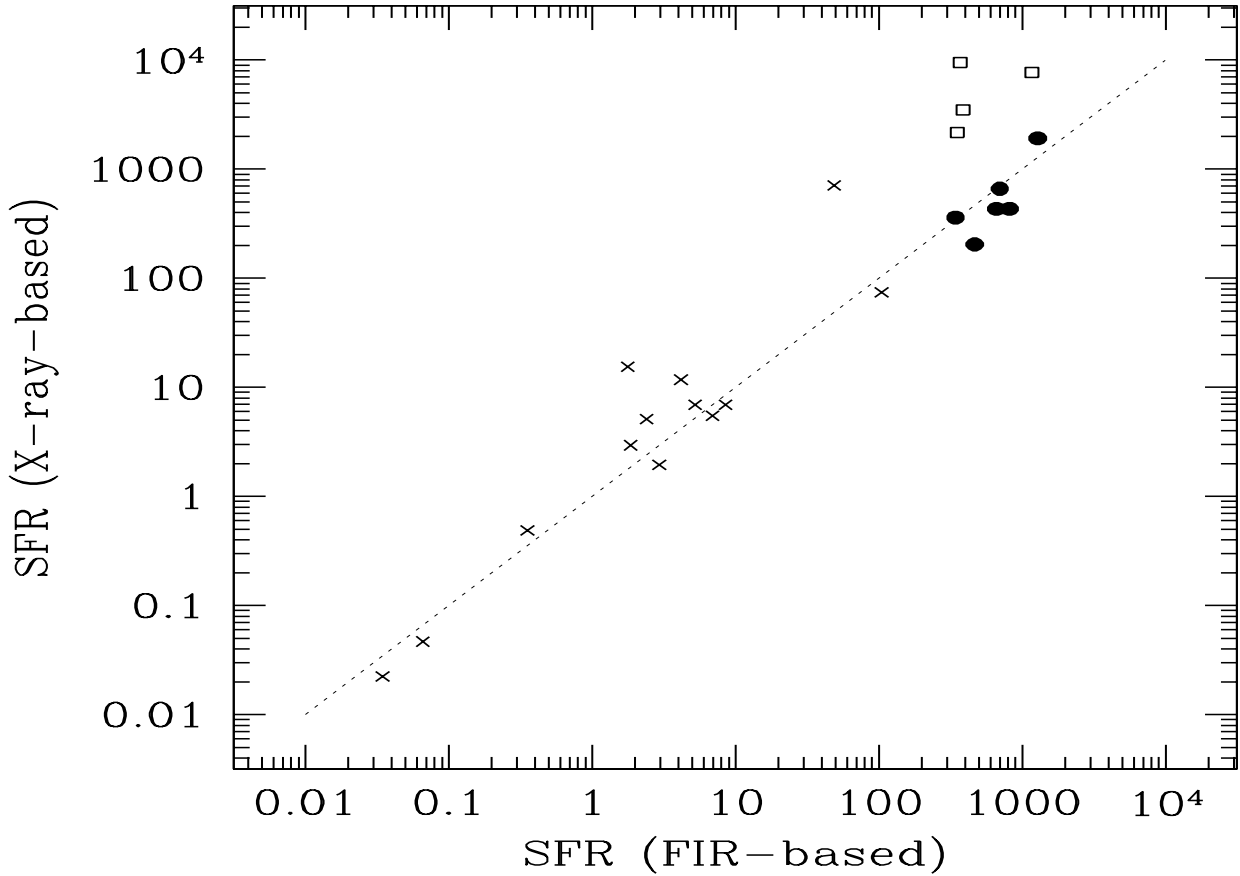,width=01.3\textwidth,height=0.99\textwidth}
\caption{-
The SFR estimated from X-ray emission versus that estimated from FIR emission. 
Empty squares and filled dots represent, respectively, AGN-dominated and 
non-AGN-dominated ULIRGs (from our sample), while crosses denote SB galaxies 
(from the literature). We used a representative SB sample (see section 5.2), 
for which: 
{\it a)} distances and FIR luminosities come from Shapley et al. 2001 
(M82, M83, NGC 253, NGC 1569, NGC 3079, NGC 3310, NGC 3628, NGC 3690, NGC 4631), 
Dahlem et al. 1998 (NGC 55, NGC 4666), Della Ceca et al. 1999 (NGC 2146), 
and Moran et al. 1999 (NGC 3256); and 
{\it b)} 2-10 keV fluxes come from Dahlem et al. 1998 (M82, NGC 55, NGC 253, NGC 
3079, NGC 3628, NGC 4631), Okada et al. 1997 (M83), Della Ceca et al. 1996 
(NGC 1569), Della Ceca et al. 1999 (NGC 2146), Moran et al. 1999 (NGC 3256), 
Zezas et al. 1998 (NGC 3310, NGC 3690), and Persic et al. 2003b (NGC 4666).
The dotted line marks the relation SFR$_{\rm X-ray}$ = SFR$_{\rm FIR}$.
The source with SFR$_{X}\simeq$750 and SFR$_{FIR}\simeq$50$M_\odot/yr$ 
corresponds to NGC 3690 (ARP 299), which was found by Della Ceca et al. (2002) to
contain an absorbed AGN.
}
\label{binaries}
\end{figure*}

\section{DISCUSSION}

Altogether, in spite of the limited statistics of the sample, there are some
remarkable regularities emerging from the spectral analysis of the XMM-{\it Newton} data. 

The first one is that hot thermal plasma emission, likely a SB signature, 
is present in all objects of the sample. We find, however, a remarkable lack 
of correlation between thermal emission and far-IR luminosity 
(which is supposed to be a good indicator of the rate of star formation in the galaxy).
Our conclusion is that, although this plasma emission should be traced back 
to a galactic wind triggered by young and exploding stars, the process may be
ruled by various additional parameters, like the density and pressure of the
surrounding ISM. 
According to the study by Strickland \& Stevens (2000), the soft X-ray emission 
from SB-driven galactic winds comes from a low filling-factor 
($\leq 2$\%) gas, which contains only a small fraction ($\leq 10$\%) of the mass
and energy of the wind. Soft X-ray observations therefore do not probe the gas that
contains the bulk of the energy, mass or metals in the outflow, while the bulk of
the hot plasma radiates and cools on timescales longer than that of the starburst 
evolution. 

To test the possible effect of the long cooling time, we have attempted
to compare the thermal X-ray emissivity with the starburst ages. To this end, 
since all optical counterparts of our sources show double or multiple nuclei
(supposed in a merging system), we have adopted as a rough estimate of the merger stage
the separation of the nuclear components, and investigated possible relations between
the ratio of the thermal X-ray to IR luminosities against nuclear separation.
We found essentially
no correlations in these plots, which suggests that various parameters in addition
to the star-formation rate (SFR), like the pressure of the surrounding medium,
should affect the thermal plasma emission.
In any case, the soft thermal emission is a quite poor tracer of the ongoing SFR.

On the contrary, the PL emission fitting the hard X-ray excesses in the spectra 
of our SB-dominated ULIRGs seems well correlated with L$_{FIR}$.
We understand this as an effect of both quantities being closely linked to
the newly formed stellar populations, in one case (L$_{FIR}$) due to dust
reprocessing of the UV flux by young massive stars, in the other (L$_{PL}$)
to the number of HMXBs which are a subset of the young stellar population (see below).

\subsection{Stellar contribution to the X-ray emission}

We now check the hypothesis that, in ULIRGs not dominated by an AGN, the hard 
X-ray 2-10 keV emission may be mainly due to luminous ($L_{2-10} >10^{37}$ erg s$^{-1}$) 
HMXBs. To this aim, we compare the rate of star-formation (SFR) estimated from 
the X-ray flux with that traced by the FIR, under the assumption
that both the X-ray emitting HMXBs and the OB stars heating the dust responsible for
the FIR flux are short-lived, in which case they can both be used as indicators 
of the ongoing SFR. Our definition of the SFR assumes a standard Salpeter stellar 
IMF between 0.1 and 100 M$_\odot$.

We proceed as follows. Assuming a mean HMXB luminosity of 
L$_{2-10}=5 \times 10^{37}$ erg s$^{-1}$ (e.g., White et al. 1983), we first estimate 
the number of HMXBs from the 2-10 keV luminosity for each sample object.
We consider that our Galaxy, hosting $\sim 50$ bright HMXBs 
(Iben et al. 1995, and references therein), has a SFR of 
$\sim 3 \,M_\odot$ yr$^{-1}$ (e.g., Matteucci 2002). From the number of 
HMXBs we then estimate the corresponding SFR$_{\rm X-ray}$. 

Then from the FIR luminosity, we calculate (Kennicutt 1998)
$$ SFR_{\rm FIR} \simeq L_{\rm FIR}/(2.2\times 10^{43} erg\ s^{-1}) M_\odot yr^{-1}.$$
We plot the two independent estimates of the SFR in Fig. \ref{binaries}: this shows
that the two are in quite good agreement for these non-AGN-dominated objects (filled 
circles), whereas the SFR$_{\rm X-ray}$ values are clearly in excess for the 
AGN-dominated sources (empty squares). Given the linear relation between SFR,
far-IR and X-ray luminosities, 
this is obviously nothing else than the $L_{\rm FIR}-L_{\rm 2-10\, keV}$ plot of 
Fig. 6d, recast in different units.

We have further checked this X-ray to FIR relationship on a sample of well-known 
moderate-to-low luminosity local SB galaxies, shown in Fig.\ref{binaries} as crosses
(see figure's caption for references). 
While comparing these with ULIRGs, and following our previous discussion in 
Sect. \ref{SB}, we bear in mind one relevant difference between the two classes: 
the star-forming activity in ULIRGs is very intense, short-lived, and dominates the
whole galaxy, so the hard X-ray flux is produced mainly by HMXBs. In 
lower SFR starburst galaxies the X-ray flux is produced by a mix of HMXBs and LMXBs, 
coming respectively from SB regions and from the underlying quiescent disk, in comparable
proportions. In Fig. \ref{binaries} we have quantified this effect by the parameter $f$, 
the fraction of 2-10 keV emission attributable to HMXBs, 
setting $f=1$ for ULIRGs and $f\simeq 0.25$ for SB's (Persic \& Rephaeli 2002).

These results confirm that, for ULIRGs without dominant AGN components,
the 2-10 keV flux from ULIRGs is a good SFR indicator:
\begin{equation}
{\rm SFR_{X-ray}}^{\rm ULIRG} \simeq { L_{2-10}\over (10^{39} erg\ s^{-1}) }
M_\odot yr^{-1}
\end{equation}
for $L_{2-10}\geq 10^{41} erg\ s^{-1}$.
For lower luminosity SB galaxies ($L_{2-10}< 10^{41} erg\ s^{-1}$), 
the relation of the X-ray flux to the SFR is less straightforward, due to the 
contribution of long-lived LMXBs. In this case the relation becomes
\begin{equation}
{\rm SFR_{X-ray}}^{\rm SB} \simeq ({f \over 0.25})
{ L_{2-10}\over (4 \times 10^{39} erg\ s^{-1}) }
M_\odot yr^{-1}.
\end{equation}
These results are in fair agreement with those found by Ranalli et al. (2002) and 
Gilfanov et al. (2003), if we consider the different definitions of the SFR. 
Note that the only SB galaxy lying far out the one-to-one relation
in Fig. \ref{binaries} is the crossed symbol with SFR$_{X-ray}\sim 750\ M_\odot yr^{-1}$
corresponding to the galaxy Arp 299, which was recently proven to host a luminous
obscured AGN (Della Ceca et al. 2002).

Finally, our XMM-{\it Newton} data below $\sim 0.5$ keV have in all
cases a very low S/N, which does not allow us to constrain the origin of this
part of the ULIRG spectrum. In some instances (IRAS12112+0305 and IRAS 22491-1808)
our formal fit implies a dominance of X-ray binary emission at such low
energies, which may not be physical due to the fact that binary spectra 
typically show photoelectric absorption. To address this problem with more
elaborated model spectra would need higher spectral resolution and S/N data.

\subsection{The AGN contribution to the ULIRG phenomenon}

Altogether, we find in 3 of the 10 sample sources (Mkn 231, IRAS 19254-7245 and 
IRAS 20551-4250) various independent evidences 
for the presence of an absorbed AGN dominating the XMM-{\it Newton} X-ray spectrum.
They show, in particular, values of the 2-10 keV luminosity 
(after absorption correction) quite in excess compared with the thermal plasma 
luminosities.
This evidence is confirmed in all three objects by X-ray spectral features like a
strong Fe-K 6.4 keV line and a very flat or inverted hard X-ray spectrum 
indicative of high-column density (N$_H>10^{22}\ cm^{-2}$) circum-nuclear material.

The source IRAS 23128-5919 displays a hard spectrum and large X-ray luminosity
which suggest an AGN contribution.
Both IRAS 23128-5919 and IRAS 20100-4156 display intermediate values of the parameters
in Fig. \ref{fig_comp} between the AGN and SB-dominated sources, which might 
indicate that an AGN contribution may be present.

It should be noticed that, among X-ray AGNs, IRAS 19254-7245, IRAS 20551-4250
and IRAS 23128-5919 did not show any AGN signatures 
from spectroscopy of IR coronal lines (Genzel et al. 1998).
Consider however that, as shown 
in Fig. \ref{fig_comp}e, the X-ray luminosity is only a tiny fraction ($\sim 0.01$\% and
$\sim 0.1$\% for SBs and AGNs, respectively) of the bolometric one. Then even for
{\sl bona-fide} X-ray AGNs, the IR spectrum may well be dominated by the starburst,
hence explaining the results of the IR spectroscopy.
Although the lack of AGN signatures in the IR spectra of these sources might alternatively
be explained as dust obscuration up to 30 $\mu$m, the evidence for starburst components 
in the X-ray spectra suggests that indeed an important fraction of the bolometric IR 
flux is likely of stellar origin (see previous Sect. and Fig. \ref{binaries}).

Though limited by the small statistics of our ULIRG sample, our conclusion is that
for a majority of these sources the high rate of star formation
indicated by the large far-IR luminosity can account for most of the X-ray emission.
The X-ray spectra of roughly half of the ULIRGs show evidence of AGN contributions
on top of an, always present, starburst component.

\section{CONCLUSION}  
    
We have devoted a large observing program with XMM-{\it Newton} to survey the hard
X-ray properties of a complete and representative sample of Ultra-Luminous IR 
Galaxies, as a way to probe deeply into the heavily extinguished cores of this
still physically elusive class of sources.

All the 10 observed ULIRGs have been detected with high statistical significance 
by XMM-{\it Newton}, but their 2-10 keV fluxes turned out to be rather faint on average.

The X-ray emission appears to be extended on a scale of $\sim$30 kpc for
Mkn 231 and IRAS 19254-7245, possibly evidence of galactic superwinds. 
In these same sources, in IRAS 20551-4250 and IRAS 23128-5919 we 
find evidence for the presence of hidden AGNs, while a 
minor AGN contribution may be suspected also in IRAS 20100-4156.
A strong Fe-K line (EW$\sim 2$ keV) in the X-ray spectrum of IRAS 19254-7245 
and a much weaker line in Mkn 231 (EW$\sim 0.2$ keV) are also detected, 
suggestive of deeply buried type-II quasars (Braito et al. 2003a,b).
For the other sources, hence for roughly half of the ULIRG sample, 
the X-ray luminosities and 
spectral shapes are mostly consistent with hot thermal plasma and X-ray binary 
emissions of mainly starburst origin.

We have analysed the XMM-{\it Newton} spectra in terms of various physical components,
that is thermal plasma emission and hard X-ray power-laws due to either X-ray
binaries or true AGNs. We have found interesting regularities in the source
parameters.  Thermal plasma emission, the signature of a starburst component
and dominating the spectra between 0.5 and 1 keV, is present in all the sample objects. 
The alternative interpretation of this soft X-ray component as due to a
reflected emission by AGN-photoionized gas is mostly excluded by our analysis.

This thermal emission is quite unrelated with the far-IR luminosity (a tracer of the
ongoing star formation rate), from which we infer that other parameters should 
determine it in addition to the rate of star formation.

On the contrary, the X-ray binary power-law emission fitting the hard X-ray 
component in the starburst-dominated ULIRGs is correlated with L$_{FIR}$,
both quantities tracing the number of young stars in the galaxy and measuring the
ongoing SFR.

We fully confirm with these data the composite nature of ULIRGs as a class,
with indications for a predominance of the starburst over the AGN phenomenon even
when observed in hard X-rays.

\section*{Acknowledgments}               
We thank the referee, dr. K. Iwasawa, for his careful reading of the paper 
and very useful comments.
This work received financial support from ASI (I/R/037/01 and I/R/062/02) under the 
project ``Cosmologia Osservativa con XMM-{\it Newton}" 
and support from the Italian Ministry of University 
and Scientific and Technological Research (MURST) through grants Cofin $00-02-004$. 
PS acknowledges financial support by the Italian {\it Consorzio  
Nazionale per l'Astronomia e l'Astrofisica} (CNAA).

\end{document}